\documentclass[aps,reprint,superscriptaddress,amsmath,citeautoscript,longbibliography]{revtex4-1}
\usepackage{graphicx}

\usepackage[breaklinks=true]{hyperref}
\usepackage[all]{hypcap}
\hypersetup{
	colorlinks = true,
	linkcolor = {blue},
	citecolor = {blue},
	urlcolor = {blue},
}

\usepackage{multirow}

\usepackage{tabularx}
\newcolumntype{Y}{>{\centering\arraybackslash}X}

\AtBeginDocument{%
    \newwrite\bibnotes
    \def\bibnotesext{Notes.bib}
    \immediate\openout\bibnotes=\jobname\bibnotesext
    \immediate\write\bibnotes{@CONTROL{REVTEX41Control}}
    \immediate\write\bibnotes{@CONTROL{%
    apsrev41Control,author="08",editor="1",pages="1",title="0",year="1"}}
     \if@filesw
     \immediate\write\@auxout{\string\citation{apsrev41Control}}%
    \fi
}%

\begin{document}

\title{Slip competition and rotation suppression in tantalum and copper during dynamic uniaxial compression}

\author{P.~G.~Heighway}
\email{patrick.heighway@physics.ox.ac.uk}
\affiliation{Department of Physics, Clarendon Laboratory, University of Oxford, Parks Road, Oxford, OX1 3PU, United Kingdom}
\author{J.~S.~Wark}
\affiliation{Department of Physics, Clarendon Laboratory, University of Oxford, Parks Road, Oxford, OX1 3PU, United Kingdom}

\date{\today}

\begin{abstract}
When compressed, a metallic specimen will generally experience changes to its crystallographic texture due to plasticity-induced rotation. Ultrafast x-ray diffraction techniques make it possible to measure rotation of this kind in targets dynamically compressed over nanosecond timescales to the kind of pressures ordinarily encountered in planetary interiors. The axis and the extent of the local rotation can provide hints as to the combination of plasticity mechanisms activated by the rapid uniaxial compression, thus providing valuable information about the underlying dislocation kinetics operative during extreme loading conditions. We present large-scale molecular dynamics simulations of shock-induced lattice rotation in three model crystals whose behavior has previously been characterized in dynamic-compression experiments: tantalum shocked along its $[101]$ direction, and copper shocked along either $[001]$ or $[111]$. We find that, in all three cases, the texture changes predicted by the simulations are consistent with those measured experimentally using \emph{in situ} x-ray diffraction. We show that while tantalum loaded along $[101]$ and copper loaded along $[001]$ both show pronounced rotation due to asymmetric multiple slip, the orientation of copper shocked along $[111]$ is predicted to be stabilized by opposing rotations arising from competing, symmetrically equivalent slip systems.
\end{abstract}

\maketitle

\section{\label{sec:introduction} Introduction}

When a malleable metal is compressed by stresses greater than its elastic limit, it will deform permanently via the motion of dislocations. Because the variety of both the dislocations and the atomic planes on which they move is limited, plastic deformation mediated in this way is often highly anisotropic. The mismatch between the plastic deformation and the imposed deformation is generally compensated for by local rotation of the metal’s crystal structure, which, at the macroscopic level, translates into evolution of its crystallographic texture. By carefully analyzing this evolution with the aid of electron backscatter diffraction (EBSD) or x-ray diffraction (XRD) techniques, it is possible in principle to infer which plasticity mechanisms are operative in the plastically deforming sample \cite{Margulies2001,Basinski2004,Florando2006,Winther2008,Chen2013,Oddershede2015,Hemery2019}. This information can in turn constrain the microscopic properties of the underlying dislocations (such as Peierls barriers, nucleation rates, and interactions strengths) that govern the metal’s plastic behavior at its most fundamental level.

There is now considerable interest in performing this kind of texture analysis at the extreme pressures made accessible by dynamic compression via laser ablation. The past few decades have seen a proliferation of high-power, long-pulse laser facilities \cite{Danson2019}, at which one can rapidly load matter to thousands or even millions of atmospheres in a shock- or ramp-compressed manner, and at a shot repetition rate that promises to reach the hertz level in the next few years \cite{Prencipe2017,Mason2018}. These platforms are often equipped to perform simultaneous ultrafast x-ray diffraction [using laser-plasma backlighters, synchrotron radiation, or emission from a hard x-ray free-electron laser (XFEL)], which can capture time-resolved images of the target’s crystal structure in its transient high-density state. Facilities like these have made possible the small but highly informative set of studies in recent years that have elucidated the texture evolution of metallic specimens under extreme loading conditions \cite{Turneaure2009,Suggit2012,Milathianaki2013,Wehrenberg2017,Turneaure2018,Sliwa2018}.

One such study that has proved particularly fruitful is that of Wehrenberg \emph{et.~al.}\ \cite{Wehrenberg2017}. In an experiment conducted at the Matter in Extreme Conditions (MEC) endstation of the Linac Coherent Light Source (LCLS), the authors obtained femtosecond x-ray diffraction patterns from textured polycrystalline tantalum foils shock-compressed along $[101]$ to over 200 GPa (two megabar). This data allowed them to deduce the crystallographic direction about which each grain locally rotated, and hence to identify the two dominant slip or twin systems of the types $[111](1\bar{2}1)$ and $[1\bar{1}1](121)$ activated by the compression process. Using this information, the measured variation of the rotation with shock pressure could be compared with the predictions of a simple kinematic model derived from the Schmid treatment of plasticity-induced rotation \cite{Schmid1926,Schmid1935}. While the model was reasonably successful -- particularly given its simplicity -- it consistently overestimated the degree of rotation by several percent, with the discrepancy becoming especially pronounced at lower shock pressures ($\lesssim$ 50 GPa).

This disparity remained largely unresolved until the recent work of Avraam \emph{et.~al.}\ \cite{Avraam2021}. In this computational study, the authors conducted finite-element simulations of the above-mentioned experiment, using a dislocation-based crystal plasticity (CP) model to treat the tantalum crystal’s constitutive response. This model, whose structure was similar in many ways to the Livermore multiscale model (LMS) model \cite{Barton2011} but was augmented by the inclusion of a dislocation-nucleation term, successfully reproduced the shock-induced rotation up to pressures of at least 150~GPa. The model further allowed the authors to distinguish two regimes of markedly different plastic behavior, separated by a transition pressure of 26~GPa: above this threshold, a confluence of dislocation-kinetics effects was found to concentrate slip almost exclusively onto just one of the two dominant slip systems, causing maximal rotation of the local crystal structure; by contrast, activity was distributed far more evenly between these two opposing slip systems below 26~GPa, leading to the same abrupt drop in lattice rotation observed by Wehrenberg \emph{et.~al.} at lower shock pressures.

Not only, then, did this study provide the first model capable of accounting precisely for the tantalum foils’ texture evolution in terms of their underlying dislocation kinetics, but it also underscored the importance of competition between different plasticity mechanisms in determining the extent of the final rotation. In this instance, the `contrast' between the activities on the opposing $[111](1\bar{2}1)$ and $[1\bar{1}1](121)$ systems controlled the stability of tantalum's $[101]$ direction against shock compression. Stabilization of certain tension or compression axes due to equal activation of symmetrically equivalent slip systems is a phenomenon widely reported from traditional materials testing experiments \cite{Nakada1964,Ramaswami1965,Keh1965,Vorbrugg1971,Takeuchi1975,Miura1977,Franciosi1982} and, more recently, ultra-large-scale atomistic simulations \cite{Ruiz2021}. We submit that inter-slip competition of this kind is even more important (and, in fact, almost inevitable) in a dynamic compression context: the lateral confinement conditions that prevent dynamically compressed matter from expanding in the transverse directions during the timescale of the experiment will usually require that more than one plasticity mechanism become active in order to relieve completely the shear stresses accumulated during compression. In other words, multiple slip, rather than single slip, is likely the norm under shock-compression conditions. The relative orientation of the active plasticity mechanisms will govern how their rotations combine, and, crucially, whether they do so in such a way as to suppress rotation and thus stabilize the compression axis. An understanding of these effects will become vital as interest in high-pressure crystal-plasticity studies grows, and we therefore see this as an opportune moment for a dedicated investigation of this physics.

In this study, we present the results of a classical molecular dynamics (MD) simulation campaign designed to explore shock-induced texture evolution in three model crystals for which experimental data already exist: bcc tantalum compressed along [101], studied in the above-mentioned experiment of Wehrenberg \emph{et.~al.}\ \cite{Wehrenberg2017}; fcc copper compressed along [001], which was measured in an experiment conducted at the Jupiter Laser Facility (JLF) by Suggit \emph{et.~al.}\ \cite{Suggit2012}; and copper compressed along [111], which was investigated by Milathianaki and coworkers \cite{Milathianaki2013} at the Coherent X-ray Imaging (CXI) instrument at LCLS. Our overarching goals are to characterize the texture evolution precipitated by the shock (if any), and to explain this evolution in terms of the plasticity mechanisms responsible. We will show that, consistent with their corresponding experiments, the simulations predict that the orientation of both tantalum shocked along $[101]$ and copper shocked along $[001]$ is unstable, while the orientation of copper shocked along $[111]$ remains largely unchanged. We will explain how the rotation (or lack thereof) in each case comes about via the combination of plasticity mechanisms driven by the shock-induced shear stresses.

The paper is laid out as follows. In Sec.~\ref{sec:methodology}, we describe the setup of the large-scale molecular dynamics simulations used to model shock-induced rotation, and outline the techniques used to characterize both the active plasticity mechanisms and the attendant rotation. We then present the results of our simulations in Sec.~\ref{sec:results}; tantalum shocked along $[101]$, copper along $[001]$, and copper along $[111]$ are described in Secs.~\ref{sec:101Ta}, Sec.~\ref{sec:001Cu}, and Sec.~\ref{sec:111Cu}, respectively. We then provide a brief discussion in Sec.~\ref{sec:discussion}, before concluding in Sec.~\ref{sec:conclusion}.

\section{\label{sec:methodology} Methodology}

\subsection{\label{sec:setup} Simulation setup}

To study plasticity-induced rotation in tantalum and copper under dynamic loading conditions, we perform classical molecular dynamics (MD) simulations with the open-source code \textsc{lammps} \cite{Plimpton1995}. Atomic interactions in tantalum are modeled using the Ravelo Ta1 potential \cite{Ravelo2013}, which successfully reproduces the equation of state, elastic constants, Hugoniot particle velocities \cite{Mitchell1981}, and several other high-pressure properties \cite{Cynn1999,Dewaele2004,Dewaele2010} of tantalum in the megabar regime, and has come to be widely used in atomistic studies of tantalum under extreme loading conditions \cite{Ruestes2014,Lu2015,Tang2017,Hahn2017,Wehrenberg2017,Sliwa2018,Heighway2019,Heighway2019b}. For copper, we use the well-established Mishin potential \cite{Mishin2001}, which has been used extensively in simulations of shock-loaded copper \cite{Bringa2004,Davila2005,Bringa2006,Shehadeh2006,Cao2007,Murphy2010}, and is thus well-characterized in the high-pressure regime of interest here. Both potentials have the advantage of being of the embedded-atom-method (EAM) type, making them relatively computationally inexpensive, and therefore well-suited to performing large-scale simulations.

The targets simulated herein are all defect-free single crystals with dimensions of 0.5 \textmu m along the compression direction $(z)$ and at least 100 unit cells in the two directions transverse to the shock $(x,y)$. The crystals are subjected to periodic boundary conditions (PBCs) on those of their faces perpendicular to the shock, and are left aperiodic along $z$ in order that they can be dynamically compressed by a piston (as explained shortly). Prior to compression, the crystals are thermalized for 5~ps under constant-NVE conditions until their temperature has stabilized at an ambient value of 300~K.

The crystals are dynamically compressed by means of a piston, a block of atoms of effectively infinite mass driven into the target at fixed speed $U_P$ along the $z$-direction. The piston velocity is ramped up over a short but finite time (one picosecond) to avoid overshocking the material in close proximity to the piston. All simulations are run under a microcanonical ensemble with a 1-fs timestep until the velocity of the rear surface of the crystals exceeds a certain fraction of the particle velocity $U_P$, at which point shock breakout is determined to be imminent, and the simulation is automatically stopped.

\subsection{\label{sec:characterisation_rotation} Rotation characterization}

To measure the reorientation suffered by our simulated targets' underlying crystal structure following shock compression, we generate synthetic pole plots that allow us to see at a glance the distribution of directions of select crystal planes. To do so, we first identify every atom in the crystal with a crystalline environment (according to adaptive common neighbor analysis (aCNA) \cite{Stukowski2012b}) and calculate its local elastic deformation gradient $F^e$, the matrix mapping the ideal displacements of its nearest neighbors onto the actual displacements of these neighbors on the timestep of interest. For fcc copper, we include in this calculation each atom's 12 nearest neighbors (from which it is separated by $\frac{1}{2}\langle101\rangle$), while for bcc tantalum, we use its 14 nearest and next-nearest neighbors (displaced by either $\frac{1}{2}\langle111\rangle$ or $\langle001\rangle$). The value of $F^e$ assigned to each nondefective atom is simply the average of every possible mapping calculable from a choice of three of its neighbors' displacements.

Once the deformation state of every well-defined unit cell has been obtained in the manner above, we select a crystal plane of interest -- often one initially aligned with one of the coordinate axes -- and calculate its new local surface normal $\mathbf{N}$ via the transformation
    \begin{equation}
        \label{eq:NtoN}
        \mathbf{N}\ \to\ [(F^e)^T]^{-1}\,\mathbf{N}.
    \end{equation}
This calculation is performed for every atom in the region of interest for which $F^e$ can be defined. The distribution of orientations of $\mathbf{N}$ is then plotted on a stereogram, where the normalized crystal plane normal $\hat{\mathbf{N}}=(\hat{N}_x,\hat{N}_y,\hat{N}_z)$ is mapped onto the point
    \begin{equation}
        \label{eq:mapping}
        (X,Y) = \left(\frac{\hat{N}_x}{1 + \hat{N}_z},\frac{\hat{N}_y}{1 + \hat{N}_z}\right).
    \end{equation}
According to this mapping, a crystal plane aligned with the compression direction $\mathbf{e}_z$ appears at the origin of the stereogram, while any plane whose normal is found perpendicular to $\mathbf{e}_z$ will lie on the unit circle. Since single crystals are used throughout this study, the distribution of local orientations is, even after shock, extremely narrow, meaning we will only ever need to focus upon very small regions of these pole plots.

We note in passing that there is a material difference between calculating how the crystallographic planes transform and how the crystallographic \emph{directions} transform. If the texture changes effected by the shock really were a pure rotation, the two would transform identically (i.e.\ $[(F^e)^T]^{-1} = F^e$). However, the highly constrained nature of uniaxial compression means this will generally not be the case -- some degree of distortion of the unit cell often takes place (as we shall see). In other words, we should not assume that the crystallographic vectors $[xyz]$ and $(xyz)$ remain parallel after shock compression. We opt to calculate pole plots for the crystal planes rather than crystal directions, as it is ultimately the former that one measures directly with x-ray diffraction.

\subsection{\label{sec:characterisation_plasticity} Plasticity characterization}

To identify the plasticity mechanisms responsible for the rotation of the crystal structure, we use two complementary characterization techniques.

To detect full dislocation slip, we employ a form of slip vector analysis (SVA) \cite{Zimmerman2001}, a technique that identifies atoms whose original neighbors have been permanently displaced from their original positions in their unit cell. The displacement vector can be identified with the Burgers vector of the dislocation responsible for the shear motion, while the surface formed by many such atoms identified using SVA reveals the slip plane. Here, we use the MD analysis and visualization software \textsc{ovito} \cite{Stukowski2010} to shrink-wrap slipped atoms with a surface mesh, the orientation of whose faces allows us to index automatically the slip plane to which each atom belongs. We then use \textsc{ovito} to visualize atoms according to the slip system each forms part of. A more complete description of our SVA technique is given in Ref.~\cite{Heighway2019}.

To detect stacking faults or deformation twins, we require a different technique, because the associated atomic displacements are so small that the SVA technique can become unreliable for high-temperature crystals. We therefore appeal instead to a template matching technique (TMT), which categorizes atoms by finding the template unit cell to which their own unit cell is most similar. To identify atoms within a deformation twin in tantalum, for instance, we would take the set of the nearest and next-nearest ideal neighbor vectors for the host crystal, $(\{\mathbf{B}_i\},\ i=1,...,14)$, and reflect them in the twin plane of interest with unit normal $\mathbf{n}$:
    \begin{equation}
        \label{eq:reflection}
        \mathbf{B}_i\ \to\ \mathbf{B}_i - 2\left(\mathbf{B}_i \cdot \mathbf{n}\right)\mathbf{n}.
    \end{equation}
Similar bases can be constructed for all anticipated twins. We then take each nondefective atom in turn, and pair each of its neighbor vectors with an ideal neighbor vector from each basis according to their cosine similarity. The basis which maximizes the total cosine similarity summed over every pair of neighbor vectors is selected and used to categorize the atom. The same approach can be used to identify the stacking faults that form in shock-compressed face-centered cubic (fcc) copper; the only difference is that the template bases are not reflections of the host lattice, but hexagonal close-packed (hcp) unit cells with a known orientation relationship with the host lattice. Further details are given in Ref.~\cite{Heighway2019}.

\section{\label{sec:results} Results}

\subsection{\label{sec:101Ta} [101] Tantalum}

The first of the three cases we consider is body-centered cubic (bcc) tantalum compressed along its $[101]$ direction. Tantalum has garnered considerable interest from the dynamic compression community \cite{Murr1997,Hsiung2000,Higginbotham2013,Florando2013,Tramontina2014,Comley2013,Ravelo2013,Wehrenberg2015,Wehrenberg2017,Sliwa2018,Heighway2019,Heighway2019b} in part due to its high phase stability along the Hugoniot. Under shock, tantalum is thought to retain its ambient bcc structure until it shock-melts at pressures of around 300~GPa \cite{Wehrenberg2017}. While several studies have reported evidence of a pressure-driven bcc-hexagonal phase transition in tantalum recovered from 30-70~GPa \cite{Hsiung2000,Hsiung2010,Lu2015}, the interpretation of the electron diffraction patterns used to detect it remains disputed \cite{Cayron2021} and uncorroborated by x-ray diffraction measurements conducted at similar pressures \cite{Wehrenberg2017}. The apparent persistence of the bcc phase makes tantalum an ideal platform for studying plasticity under extreme loading conditions without added complications arising from solid-solid phase transitions.

When compressed along $[101]$, tantalum exhibits a particularly rich plastic response mediated by a characteristic mixture of deformation twinning and dislocation slip. Competition between these two mechanisms was observed directly in tantalum dynamically compressed to up to 200~GPa in the previously mentioned experiment of Wehrenberg \emph{et.\ al.}\ \cite{Wehrenberg2017}. In their study, the authors were able to track the texture evolution of shock-loaded polycrystalline tantalum foils, whose constituent grains had their $[101]$ direction preferentially aligned with the shock direction, by means of ultrafast x-ray diffraction. By examining the motion (or lack thereof) of scattering peaks from the $\{101\}$ planes, the authors concluded that each grain underwent plasticity-induced rotation about its local $[10\bar{1}]$ direction. They further deduced that while the dominant means of plastic deformation switched from twinning to slip between shock pressures of 100 and 150~GPa, the measured rotation axis was identical at every pressure, and was consistent with slip or twinning of the kinds $[111](1\bar{2}1)$ and $[1\bar{1}1](121)$. These twin variants had in fact been predicted by MD simulations \cite{Ravelo2013,Tramontina2014,McGonegle2015,Lu2015} and observed in shock-recovery experiments \cite{Florando2013,Lu2015}, but this was the first direct, experimental confirmation of their forming under dynamic-compression conditions.

However, it cannot be the case that activation of the $[111](1\bar{2}1)$ and $[1\bar{1}1](121)$ slip and twin systems is the whole story: the plastic strain brought about by these systems is confined to a plane, making it impossible for these plasticity mechanisms alone to bring about the kind of quasihydrostatic, isotropic elastic strain states observed in experiment \cite{Wehrenberg2017}. The remaining shear strains must therefore be relieved either via interactions between neighboring grains (a possibility explored in Ref.~\cite{Heighway2019}), or by activity on additional slip or twin systems. The latter scenario is intriguing, because it would suggest that the crystals studied by Wehrenberg \emph{et.\ al.}\ rotated as if only $[111](1\bar{2}1)$ and $[1\bar{1}1](121)$ were active \emph{despite} other plasticity mechanisms also being operative. To our knowledge, no account of plasticity in tantalum shocked along $[101]$ has yet been given that explicitly explains how it can rotate about $[10\bar{1}]$ while also relaxing to a state of minimal shear strain, and in such a way as to respect lateral confinement conditions. It is for this reason that we believe it appropriate to revisit the experiment of Wehrenberg \emph{et.\ al.}, and to attempt to build a more complete picture of the measured plasticity-induced texture evolution with the aid of MD simulations.

Our main objective is to understand how the plasticity mechanisms activated by the shock-compression process work together to bring about the overall lattice rotation. We will therefore first establish which plasticity agents are operative (according to the Ravelo potential) in tantalum shocked along $[101]$ (which we will refer to henceforth as `$[101]$ Ta' for brevity) before trying to understand the resulting reorientation of its underlying crystal structure. To do so, we will begin by focusing on a tantalum crystal shocked to a pressure exceeding only slightly its Hugoniot elastic limit (HEL): at such a `modest' pressure, the density of the crystal defects nucleated by the shock is minimal, making what defects exist considerably easier to visualize. The initial orientation of this and all other crystals is such that their $[10\bar{1}]$, $[010]$, and $[101]$ directions are parallel to the $x$-, $y$-, and $z$-axes, respectively.

\begin{figure}[t]
    \includegraphics{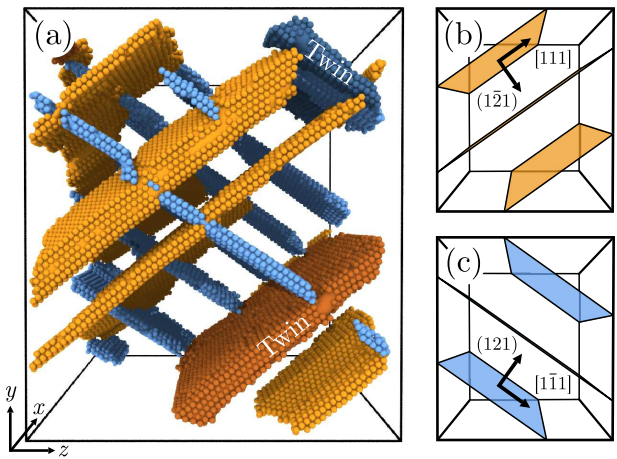}
    \caption{Primary slip systems active in bcc tantalum shock-compressed along $[101]$ to 40~GPa. (a) Visualization of atoms that have participated in slip events according to slip vector analysis (SVA); orange atoms have suffered displacement from their neighbors by $\pm\frac{1}{2}[111]$, blue atoms by $\pm\frac{1}{2}[1\bar{1}1]$. Shown also are the equivalent deformation twins, which involve displacement by $\pm\frac{1}{6}[111]$ and $\pm\frac{1}{6}[1\bar{1}1]$, respectively, identified using the template matching technique (TMT) and colored slightly darker for clarity. (b,c) Simplified depiction of the primary slip and twin systems, namely $[111](1\bar{2}1)$ (orange) and $[1\bar{1}1](121)$ (blue); these systems are referred to herein as PS1 and PS2, respectively.}
    \label{fig:101TaPrimaries}
\end{figure}

\begin{figure}[t]
    \includegraphics{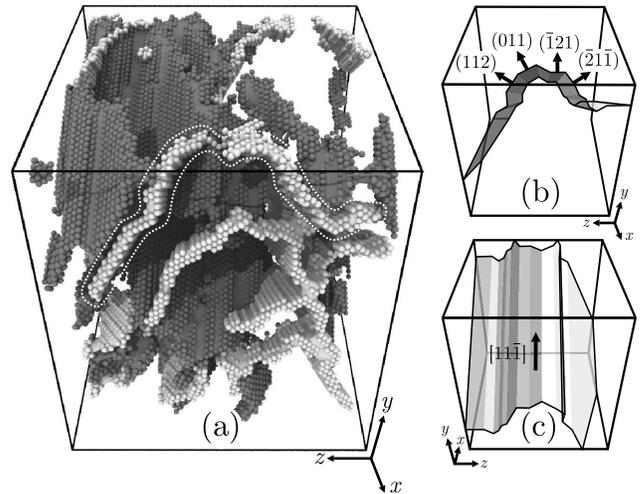}
    \caption{Secondary slip systems active in bcc tantalum shock-compressed along $[101]$ to 40~GPa. (a) Visualization of atoms that have participated in slip events according to slip vector analysis (SVA); light gray atoms have suffered displacement from their neighbors by $\pm\frac{1}{2}[11\bar{1}]$, dark gray atoms by $\pm\frac{1}{2}[\bar{1}11]$. (b,c) Simplified depiction of the slip plane structure formed by a mobile $\frac{1}{2}[11\bar{1}]$ dislocation, viewed from two different directions. Cross-slip has permitted this dislocation to move on multiple planes of the types $\{121\}$ and $\{101\}$. These systems are enumerated in Tab.~\ref{tab:101TaSystems}.}
    \label{fig:101TaSecondaries}
\end{figure}

In Figs.~\ref{fig:101TaPrimaries} and \ref{fig:101TaSecondaries}, we depict the two families of plasticity mechanisms active in $[101]$ Ta shocked to 40~GPa (the HEL for the Ravelo Ta1 potential employed here \cite{Ravelo2013}) detected using SVA and TMT. The first family, depicted in Fig.~\ref{fig:101TaPrimaries}, comprises the aforementioned slip systems and deformation twins of the types $[111](1\bar{2}1)$ and $[1\bar{1}1](121)$. Their geometry is relatively simple: for these primary systems, the slip direction $\mathbf{m}$, the slip plane normal $\mathbf{n}$, and the compression direction $\mathbf{e}_z$ all lie in the same plane, which in this instance is the $yz$-plane. The primary systems therefore cause no atomic motion in the $x$ direction; any changes in elastic strain and crystal orientation they cause are confined to the $yz$-plane. According to the conventional understanding of plasticity-induced rotation, in which the reorientation is such as to cause the active slip plane to rotate towards the compression axis, systems of the type $[111](1\bar{2}1)$ (referred to here as primary system 1, or PS1) would cause the crystal to rotate counterclockwise about $x$ when viewed as in Fig.~\ref{fig:101TaPrimaries}; the complementary systems of the type $[1\bar{1}1](121)$ (named PS2), being mirror images of PS1 in the $xz$-plane, would cause rotation in the clockwise sense instead. Note that we will not differentiate between equivalent slip and twin systems for the purposes of this study, since the sense of the rotation each causes is identical. It is to these primary plasticity mechanisms that the lattice rotation observed in the work of Wehrenberg \emph{et.\ al.}\ was ascribed \cite{Wehrenberg2017}.

There also exists a second family of deformation mechanisms (depicted in Fig.~\ref{fig:101TaSecondaries}) that consists of slip systems involving slip in either the $[11\bar{1}]$ or the $[\bar{1}11]$ directions, which sit in the $xy$-plane. These secondary systems are considerably more diverse than the primaries due to the fact that their constituent dislocations are liable to cross-slip. For illustration, we show in Figs.~\hyperref[fig:101TaSecondaries]{2(b,c)} the slip surfaces swept out by a meandering $\frac{1}{2}[11\bar{1}]$ dislocation, which can be seen moving on planes $(112)$, $(011)$, and $(\bar{1}21)$, among others. The `purpose' of these secondary systems is essentially to relieve those components of the shear stress that the primaries cannot; it is only via activity on the secondaries that the elastic strain along $x$ (i.e.\ the $[10\bar{1}]$ direction) can change under the lateral confinement conditions to which the single crystal is subjected. The change in crystallographic orientation caused by these secondary systems -- particularly those whose slip plane normals make an oblique angle with the compression direction -- is less intuitive than that caused by the primaries, but can in principle be understood using the appropriate kinematics, as will be shown. We now go on to study the rotation undergone by a representative element of material in $[101]$ Ta shocked to 40~GPa, and to attempt to reconcile this rotation with the activity of these primary and secondary systems, which are enumerated in Table.~\ref{tab:101TaSystems}.

\begin{table}[b]
\begin{tabularx}{8.5cm}{@{}YYYY@{}}
\hline
\hline
Direction $\mathbf{m}$ & Plane $\mathbf{n}$ & Mechanism(s) & Notation\\
\hline
$[111]$ & $(1\bar{2}1)$ & Slip, twins & PS1\\
\hline
$[1\bar{1}1]$ & $(121)$ & Slip, twins & PS2\\
\hline
\multirow{3}{*}[-0.5ex]{$[11\bar{1}]$} & $(\bar{1}21)$ & Slip & SS1\\
                                       & $(2\bar{1}1)$ & Slip & SS3\\
                                       & $(112)$ & Slip & SS5\\
\hline
\multirow{3}{*}[-0.5ex]{$[\bar{1}11]$} & $(12\bar{1})$ & Slip & SS2\\
                                       & $(1\bar{1}2)$ & Slip & SS4\\
                                       & $(211)$ & Slip & SS6\\
\hline
\hline
\end{tabularx}
\caption{\label{tab:101TaSystems}Enumeration of the plasticity mechanisms active in shock-compressed $[101]$ Ta.}
\end{table}

\begin{figure}[t]
    \includegraphics{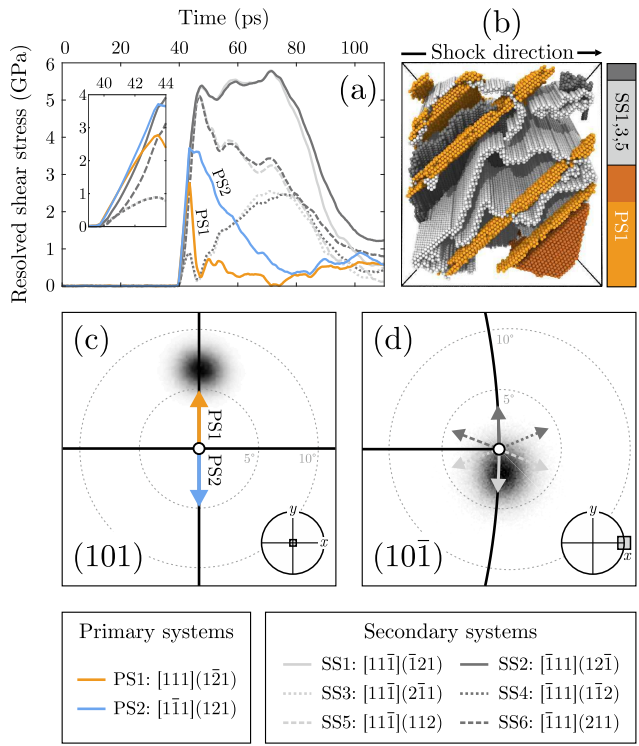}
    \caption{Behavior of a representative material element in $[101]$ Ta shock-compressed to 40~GPa. (a) Resolved shear stresses acting on primary and secondary deformation mechanisms as functions of time. (b) Visualization of operative deformation mechanisms at late times. The proportion in which the plasticity mechanisms are active is indicated by the stacked bar chart. (c,d) Pole plots showing late-time orientations of the $(101)$ and $(10\bar{1})$ planes, respectively. White circles mark the initial orientation of these planes. Arrows indicate the sense in which the orientations should change if given plasticity mechanisms operate.}
    \label{fig:101TaLagrangian}
\end{figure}

In Fig.~\ref{fig:101TaLagrangian}, we show a composite image illustrating the behavior of a material element initially situated approximately 190~nm from the piston (a distance constituting about 37.5\% of the sample's total length) during and after the shock-compression process. The cubical element in question has pre-shock dimensions of $50^3$ unit cells, and thus comprises some 250\,000 atoms. Fig.~\hyperref[fig:101TaLagrangian]{3(a)} shows the resolved shear stresses acting on its primary and secondary systems as functions of time, and in Fig.~\hyperref[fig:101TaLagrangian]{3(b)}, we show a visualization of the plasticity mechanisms that have become active within 70~ps of the shock wave passing. We also show in Figs.~\hyperref[fig:101TaLagrangian]{3(c,d)} pole plots displaying the late-time orientation distributions of the $(101)$ and $(10\bar{1})$ planes, which in our setup were originally aligned with the $z$- and $x$-directions, respectively. Together, these figures provide an overview of the typical dynamic response of $[101]$ Ta to shock compression, which, for this particular material element, proceeds as follows.

When the shock wave first passes at $t = 40$~ps, resolved shear stress accumulates most rapidly on the two primary systems, as shown in the inset of Fig.~\hyperref[fig:101TaLagrangian]{3(a)}. Within a few picoseconds, these shear stresses become great enough to trigger plasticity, and one of the two primary systems becomes active (in this instance, PS1). The ensuing plastic flow rapidly relieves the shear stress acting on PS1, and also relieves much of the stress driving PS2 (the inactive primary system), though less efficiently. There follows a period of approximately 25~ps during which sustained activity on PS1 relieves almost all of the shear stress on the primary systems (which relaxes to a limiting value 0.5~GPa), while the secondary systems, on which shear stress accumulates more slowly, remain inoperative. It is only at around $t = 70$~ps that a subset of the secondaries suddenly becomes active, and relieves the shear stress acting upon them over the following 30~ps. The shear stresses fall faster on the secondary systems involving slip along the $[11\bar{1}]$ direction (i.e.\ those labeled with odd numbers in Fig.~\ref{fig:101TaLagrangian}), suggesting they are more active than are the even-numbered secondary systems. This intuition is confirmed by the atomistic visualization in Fig.~\hyperref[fig:101TaLagrangian]{3(b)}, which reveals that many more atoms have suffered displacement by $\pm\frac{1}{2}[11\bar{1}]$ than by $\pm\frac{1}{2}[\bar{1}11]$ at late times. By $t = 110$~ps, plastic flow has all but ceased, and the combination of primary and secondary systems (namely PS1 and the secondaries involving slip along $[11\bar{1}]$) has left the crystal in a largely nondeviatoric stress state.

\begin{figure*}[t]
    \includegraphics{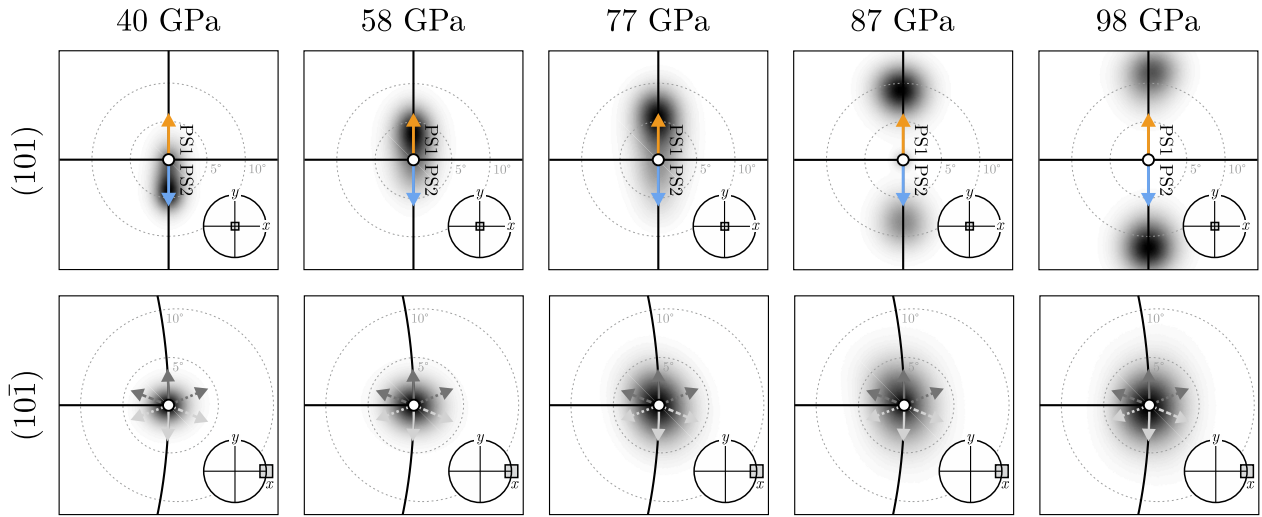}
    \caption{Pole plots showing the directions of the $(101)$ and $(10\bar{1})$ planes in $[101]$ Ta shock-compressed to between 40 and 100~GPa shortly before shock breakout. Only the piston and a sliver of uncompressed material between the shock front and the rear surface have been excluded from the calculation of these pole plots. White circles mark the initial orientation of the $(101)$ and $(10\bar{1})$ planes. Arrows indicate the sense in which the orientations should change if given plasticity mechanisms operate.}
    \label{fig:101TaPressureScan}
\end{figure*}

Having identified the plasticity mechanisms operative in this particular region of the crystal, we are now in a position to account for the overall rotation it suffers. Figs.~\hyperref[fig:101TaLagrangian]{3(c,d)} show the late-time orientations of the $(101)$ and $(10\bar{1})$ planes, respectively. They indicate that while the $(101)$ planes (which were originally parallel to the compression axis, $z$) have been deflected by approximately $7^\circ$, the $(10\bar{1})$ plane normals (originally aligned with the transverse direction, $x$) have been deflected by only a couple of degrees. We can understand this behavior by using the kinematics of slip-induced rotation to predict how each crystal plane transforms under the action of a given slip system. The central equation connecting plasticity to the concomitant change in crystal structure is the elastoplastic decomposition of the total deformation gradient $F$, which reads
    \begin{equation}
        \label{eq:elastoplastic}
        F = F^e F^p,
    \end{equation}
where $F^e$ and $F^p$ are the elastic and plastic deformation gradients, respectively. The plastic deformation attending an amount of glide $\gamma$ on the system with slip direction $\mathbf{m}$ and slip plane normal $\mathbf{n}$ is, to leading order,
    \begin{equation}
        \label{eq:plasticdef}
        F^p = I + \gamma (\mathbf{m}\otimes\mathbf{n}),
    \end{equation}
where $\otimes$ is the outer product operator. Meanwhile, the total deformation gradient for true uniaxial strain along the $z$-direction takes the form
    \begin{equation}
        \label{eq:totaldef}
        F = \text{diag}(1, 1, v),
    \end{equation}
where $v = V/V_0$ is the total volumetric compression suffered by the crystal (around 14\% for this particular simulation). Via Eqn.~(\ref{eq:elastoplastic}), operators $F$ and $F^p$ yield the elastic deformation gradient $F^e$, which describes how crystallographic directions transform; the analogous operator expressing how the orientation of crystallographic \emph{planes} changes is $[(F^e)^T]^{-1}$. Hence, the general crystal plane normal $\mathbf{N}$ transforms according to the equation
    \begin{equation}
        \label{eq:Ge}
        \mathbf{N}\ \to\ \text{diag}(1,1,v^{-1})[I + \gamma(\mathbf{n}\otimes\mathbf{m})]\,\mathbf{N}.
    \end{equation}
This equation allows us to overlay on Figs.~\hyperref[fig:101TaLagrangian]{3(c,d)} predictions of how the orientation of each plane should change under the action of any given plasticity mechanism.

When Eqn.~(\ref{eq:Ge}) is applied to the forward $(101)$ plane, we find that its orientation is sensitive only to activity on the two primary systems, which push the $(101)$ plane normal in opposite directions along the great circle joining the $y$ and $z$ axes -- this is consistent with the notion that PS2 should cause `clockwise' rotation of the crystal structure about $x$, PS1 `anticlockwise' rotation. As indicated in Fig.~\hyperref[fig:101TaLagrangian]{3(c)}, the observed reorientation of the $(101)$ plane is perfectly consistent with preferential activation of PS1, the primary system shown by SVA and TMT to be the dominant of the two for this particular material element. Analogously, when Eqn.~(\ref{eq:Ge}) is applied to the $(10\bar{1})$ plane, we find its direction depends only on the secondary systems. As shown in Fig.~\hyperref[fig:101TaLagrangian]{3(d)}, the slight reorientation of the $(10\bar{1})$ planes is consistent with preferential activation of the odd-numbered secondaries over the even-numbered ones, which, again, corroborates the story told by the visualization in Fig.~\hyperref[fig:101TaLagrangian]{3(b)}. The reorientation of the $(101)$ and $(10\bar{1})$ planes at least, then, can be explained by the combination of plasticity mechanisms active at this particular shock pressure.

As we move to higher shock pressures, the response of $[101]$ Ta remains qualitatively similar: the forward $(101)$ plane is deflected by an increasing amount around the $x$-axis, while the transverse $(10\bar{1})$ plane remains largely static -- indeed, when pole plots are generated from the entire thickness of compressed crystal, as in Fig.~\ref{fig:101TaPressureScan}, we observe that the center of mass of the $(10\bar{1})$ distribution stays resolutely aligned with the $x$-direction up to shock pressures of 100~GPa. This behavior reveals a crucial difference between the primary and the secondary systems. The fact that, on average, the $(10\bar{1})$ planes suffer no change in direction indicates that activity is distributed equally among the various secondary systems. As depicted in Figs.~\hyperref[fig:101TaLagrangian]{3(d)} and \ref{fig:101TaPressureScan}, the secondaries are found in symmetrically-equivalent pairs that drive the $(10\bar{1})$ planes in opposite directions; that $(10\bar{1})$ is practically immobile indicates that SS1 is as active as SS2, SS3 as active as SS4, and so on. By contrast, it must be the case that the symmetry between PS1 and PS2 is broken, otherwise the $(101)$ planes could suffer no net reorientation. The pole plots in Fig.~\ref{fig:101TaPressureScan} show that the crystal randomly selects one of the two primary systems to become active preferentially, sometimes favoring PS1 (and thus rotating anticlockwise), sometimes PS2 (rotating clockwise). It is this tendency of the crystal to apportion all of the plasticity to one of the two primary slip systems that renders the $[101]$ axis unstable to shock compression.

\begin{figure*}
    \includegraphics{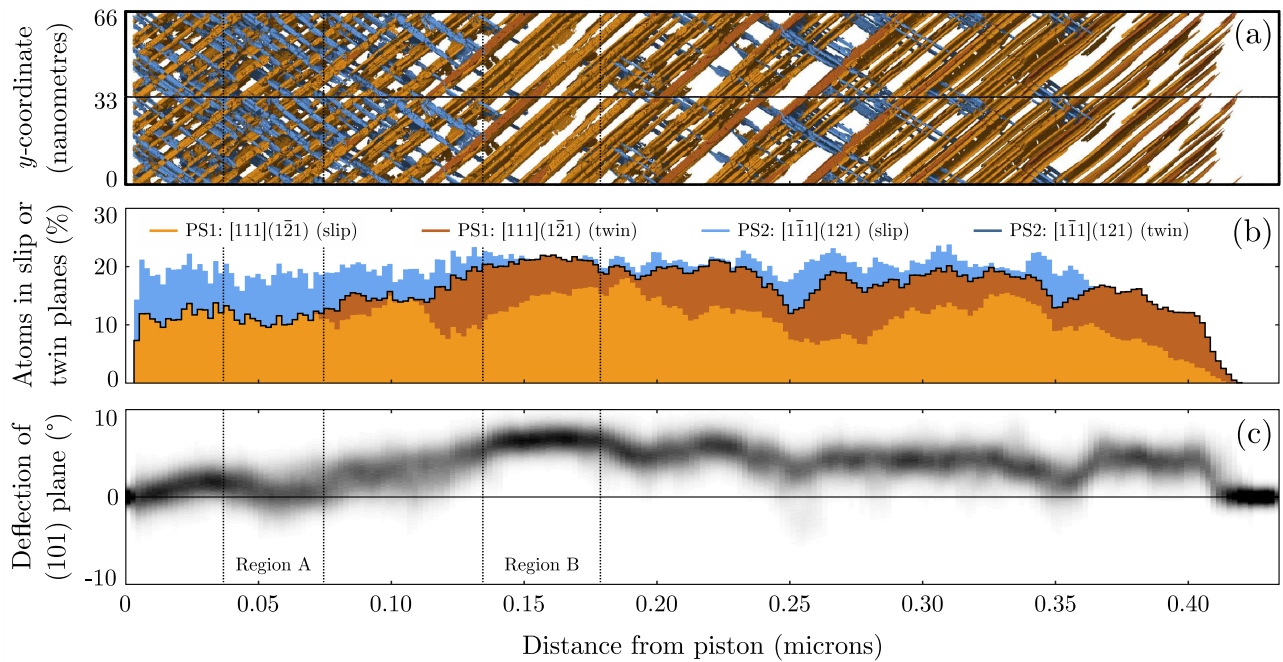}
    \caption{(a) Visualization of primary deformation mechanisms PS1 and PS2 operative in $[101]$ Ta shocked to 40~GPa, shown along the length of the crystal. The piston-driven shock wave travels from left to right. One periodic image of the crystal is shown in the $y$-direction for clarity. (b) Fraction of atoms in each spatial bin belonging to primary systems PS1 and PS2, displayed as a stacked bar chart. The black line separates contributions to PS1 (from either slip or twinning) from contributions to PS2 (which in this instance is mediated exclusively by slip). (c) Distribution of deflection angles of the $(101)$ plane about the $x$-axis as a function of distance along the crystal. Regions A and B suffer close to minimal and maximal deflection, respectively, for this given shock pressure.}
    \label{fig:101TaProfile}
\end{figure*}

\begin{figure}[t]
    \includegraphics{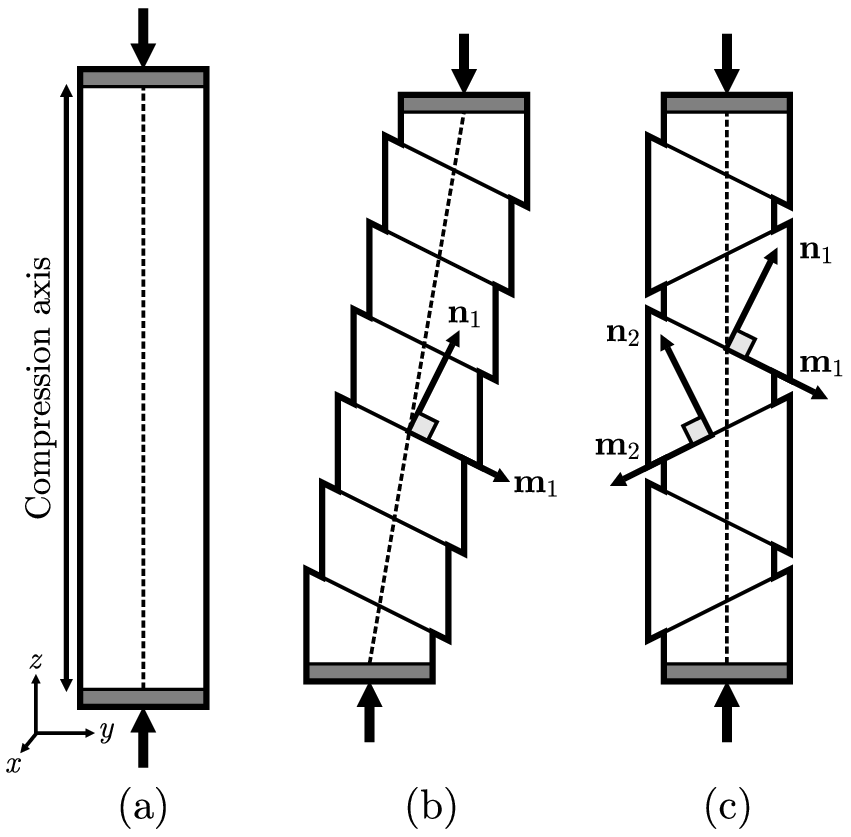}
    \caption{Schematic illustrating symmetric and asymmetric slip-mediated plastic deformation of a crystalline specimen. (a) Sample pictured before compression along $z$. (b) Sample after deforming on a single slip system with slip direction $\mathbf{m}_1$ and slip plane normal $\mathbf{n}_1$. (c) Sample after deforming on two symmetrically equivalent slip systems, the second of which has a slip direction $\mathbf{m}_2$ and a slip plane normal $\mathbf{n}_2$ that are mirror images of those of the first slip system in the $xz$-plane. Rotation would be required in the former case to ensure that the line element initially aligned with $z$ (indicated by the dashed line) retains its orientation during compression; no such reorientation is needed in the latter case.}
    \label{fig:101TaCancellation}
\end{figure}

This physics is illustrated more vividly when one examines the variation of the local rotation and plasticity along the length of the target -- it so happens that near the piston, activity on PS1 cancels almost perfectly with activity on PS2, which in turn suppresses rotation of the $(101)$ planes. In Fig.~\hyperref[fig:101TaProfile]{5(a)}, we show a visualization of the primary systems along the length of a $[101]$ Ta crystal shocked to 40~GPa shortly before breakout. We notice at once that while PS1 dominates overall, the majority of the crystal also sees a nonzero amount of activity on PS2. In fact, the two primary systems are active to almost the same extent for material within 100~nm of the piston, which is marked by densely interpenetrating slip planes of the types $(121)$ and $(1\bar{2}1)$. A raw count of the number of atoms assigned to each primary plasticity mechanism is shown in Fig.~\hyperref[fig:101TaProfile]{5(b)}, giving an idea of the level of contrast between PS1 and PS2 as a function of distance along the crystal. When this contrast profile is compared with the local deflection of the $(101)$ plane, which is plotted in Fig.~\hyperref[fig:101TaProfile]{5(c)}, we find a clear correlation between the two: the closer the activities on PS1 and PS2, the less the net rotation. The two limiting cases of this relationship are represented by the regions marked A and B in Fig.~\hyperref[fig:101TaProfile]{5(c)}: Region A, which is situated in proximity to the piston, sees almost equal activity on the two primary slip systems, and thus experiences almost no reorientation of its $(101)$ plane; Region B, which overlaps the Lagrangian material element shown in Fig.~\ref{fig:101TaLagrangian}, deforms almost exclusively on PS1, and thus experiences the greatest deflection of its $(101)$ planes possible for the given amount of volumetric compression.

The physical intuition behind rotation cancellation of this kind can be put on firm mathematical footing using the leading-order expression for the plastic deformation gradient $F^p$ arising from activity on a set of $N>1$ slip or twin systems, which reads
    \begin{equation}
        \label{eq:plasticdefpoly}
        F^p = I + \sum_{\alpha=1}^N \gamma_\alpha (\mathbf{m}_\alpha\otimes\mathbf{n}_\alpha).
    \end{equation}
Here, $\gamma_\alpha$ is the amount of glide on slip system $\alpha$ defined by its slip direction $\mathbf{m}_\alpha$ and its slip plane normal $\mathbf{n}_\alpha$. Considering the present case of primary systems PS1 and PS2, for which
    \begin{subequations}
        \label{eq:primaryvectors}
        \begin{align}
            \mathbf{m}_1 &= \begin{pmatrix}0 \\ m_y \\ m_z\end{pmatrix},\quad&
            \mathbf{m}_2 &= \begin{pmatrix}0 \\ -m_y \\ m_z\end{pmatrix}, \\
            \mathbf{n}_1 &= \begin{pmatrix}0 \\ n_y \\ n_z\end{pmatrix},\quad&
            \mathbf{n}_2 &= \begin{pmatrix}0 \\ -n_y \\ n_z\end{pmatrix},
        \end{align}
    \end{subequations}
equal activity $\gamma_1 = \gamma_2 = \gamma$ on the two competing systems would bring about a plastic deformation state
    \begin{equation}
        \label{eq:2dcancellation}
        F^p = \begin{pmatrix}1 & 0 & 0 \\ 0 & 1+2\gamma m_y n_y & 0 \\ 0 & 0 & 1+2\gamma m_z n_z\end{pmatrix}.
    \end{equation}
Symmetric activation of PS1 and PS2 thus produces a diagonal plastic strain state for which no compensatory lattice rotation is required of the target in order for it to match the total deformation gradient $F = \text{diag}(1,1,v)$. The difference between the (dominant) asymmetric deformation mode and the (much rarer) symmetric mode is illustrated in Fig.~\ref{fig:101TaCancellation}.

At this juncture, it is useful to connect the results of these simulations back to those of the experiment of Wehrenberg \emph{et.\ al.}\ \cite{Wehrenberg2017}. From their diffraction data, the authors concluded that each grain must rotate about its local $[10\bar{1}]$ direction due to asymmetric activity on what we have called the primary systems. This account is supported by our simulations: we find that the direction of the $(10\bar{1})$ planes, which are initially parallel to the proposed rotation axis, is almost invariant, while the $(101)$ planes originally aligned with the compression axis, whose orientation is controlled exclusively by these primary systems, are deflected by several degrees. That the primaries are largely responsible for the observed rotation is already well-established \cite{Wehrenberg2017,Sliwa2018,Avraam2021}; what our analysis suggests is that the crystal rotates about $[10\bar{1}]$ not because there are no other active plasticity mechanisms, but because the rotations caused by any additional mechanisms (i.e.,~the secondaries) mutually cancel. Hence, these single-crystal simulations suggest it is the asymmetric activation of the primaries combined with the symmetric activation of the secondaries that leads to the texture evolution observed by Wehrenberg \emph{et.~al.}

We should point out, of course, that the targets examined in the experiment were \emph{not} in fact single crystals, but fiber-textured polycrystals. The extent to which individual grains in the target should behave like isolated single crystals under shock-compression conditions is still largely unknown. The MD study in Ref.~\cite{Heighway2019} suggested that, up to pressures of 60~GPa at least, the aforementioned secondary slip systems could in fact be `replaced' by grain-grain interactions, which could relieve the same components of shear stress as the secondaries, but over a shorter timescale. In this scenario, there would \emph{also} be no detectable redirection of the $(10\bar{1})$ planes, because the columnar nature of the grain morphology means any rotation arising from grain-grain interactions would necessarily take place around the compression axis, which cannot be detected in a fiber-textured specimen. Hence, it is not possible (or not trivial, at least) to tell from the data whether the second stage of plastic deformation proceeds via the secondary plasticity mechanisms studied here, or by grain-grain interactions (or by some combination of the two). This question would perhaps be best-addressed by further CP studies of the kind performed by Avraam \emph{et.~al.}\ \cite{Avraam2021}: the inherently larger length-scale at which the finite-element-method framework operates means polycrystals can be simulated at just a fraction of the cost required in MD, which, when combined with its having the most faithful constitutive model of high-pressure tantalum to date, makes the Avraam model uniquely placed to explore the interplay of plasticity and intergranular interactions during dynamic compression. A study of grain-grain effects could form the basis of future work.

There is one last comment we wish to make briefly before concluding this study of $[101]$ Ta. We have examined the directions of the $(101)$ planes initially aligned with the compression axis, and of the $(10\bar{1})$ planes orthogonal to that axis. It so happened that the rotations brought about by the various plasticity mechanisms neatly decoupled, such that $(101)$ is only sensitive to the primaries, $(10\bar{1})$ to the secondaries. This is the exception rather than the rule -- in general, the direction of an arbitrarily chosen crystal plane will depend upon \emph{all} of the available plasticity mechanisms, and not in a way that is immediately intuitive. To illustrate this concept, we will show how the primaries and secondaries influence the direction of the $(110)$ plane, which originally makes an angle of $60^\circ$ with the compression axis -- it was in fact their diffraction peaks, rather than those of the $(101)$ planes, that Wehrenberg \emph{et.~al.} measured, making their behavior of more direct relevance to their experiment.

\begin{figure}[b]
    \includegraphics{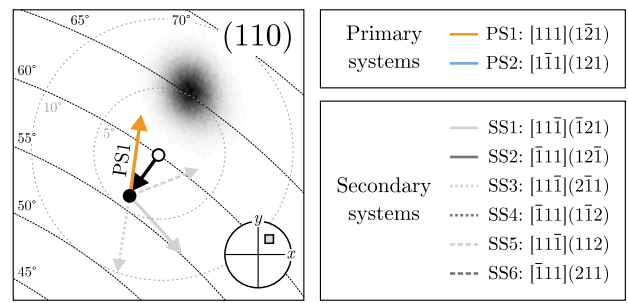}
    \caption{Pole plots showing late-time orientations of the $(110)$ planes in $[101]$ Ta shocked to 40~GPa. White circles mark the initial orientation of this plane. Black circles mark its orientation after uniaxial compression, but before plasticity. Arrows indicate the sense in which the orientation should change if given plasticity mechanisms operate. The lengths of the arrows are proportional to the amount of reorientation resulting from a fixed amount of glide $\gamma$.}
    \label{fig:101Ta60Peak}
\end{figure}

In Fig.~\ref{fig:101Ta60Peak}, we show a pole plot for the $(110)$ plane, generated from the same material element studied in Fig.~\ref{fig:101TaLagrangian}. We see at once that these planes also rotate away from the compression direction, in a manner consistent with preferential activation of PS1. However, the set of mechanisms responsible for dictating the final direction of $(110)$ differs from that controlling $(101)$. First, we notice that the total uniaxial compression itself shifts the $(110)$ plane normal by some $4^\circ$ directly towards the compression axis, even before any plasticity takes place. Purely elastic deflection of planes making an oblique angle with the compression axis can in fact be measured experimentally \cite{Milathianaki2013} and used to detect elastic precursor waves. Second, we find that the direction of $(110)$ depends on not only PS1 but also a subset of the secondary plasticity mechanisms. Specifically, the secondaries involving slip along $[11\bar{1}]$ cause $(110)$ to rotate around the compression axis -- only this can explain why the center of the orientation distribution sits so far to the `right' of the pole plot. Third, we observe that the net effect of the odd-numbered secondaries is actually to cause the $(110)$ planes to rotate slightly \emph{towards} the compression axis. This marginal dependence of the crystal's rotation state on the secondary slip systems was not considered in the original work of Wehrenberg \emph{et.\ al.}, which could account in part for the percent-level disagreement between their Schmid-type rotation model and their experimental data.

Finally, we also notice that there are a number of plasticity mechanisms to which the $(110)$ planes are \emph{not} sensitive. In particular, their direction is entirely unaffected by activity on the complementary primary system, PS2. This conclusion seems to be odds with the usual idea that compression-induced rotation is such as to cause the active slip plane to rotate towards the loading axis. How can we reconcile the facts that PS2 causes $(101)$ to rotate but leaves $(110)$ stationary, despite neither being parallel to the apparent rotation axis $[10\bar{1}]$?

The answer is that under the conditions of true uniaxial strain, a single slip system does bring about \emph{pure} rotation: it follows unavoidably from the kinematics that some degree of elastic pure shear strain (i.e.\ distortion of the unit cell) must also take place. However, there are certain planes for which the rotation and the shear cancel exactly: it may be shown using Eq.~(\ref{eq:plasticdef}) that under the action of the slip system with direction and plane normal $\mathbf{m}$ and $\mathbf{n}$, respectively, the crystal plane normal $\mathbf{N}$ is transformed according to
\begin{equation}\label{eq:NtoNalt}
\mathbf{N}\ \to\ \mathbf{N} - \gamma(\mathbf{N}\cdot\mathbf{m})\mathbf{n}.
\end{equation}
Hence, $\mathbf{N}$ is insensitive to activity on any slip (or twin) system involving shear motion orthogonal to $\mathbf{N}$ itself. This explains why the $(110)$ planes are affected by neither PS2 (which involves slip along $[1\bar{1}1]$) nor the even-numbered secondaries (which involve slip along $[\bar{1}11]$). The key point is that some of the physical intuition about texture evolution derived from traditional materials testing does not carry over into a uniaxial strain geometry; while we have used the word `rotation' extensively to describe the response of $[101]$ Ta to shock compression, we would be remiss if we did not draw attention to the fact that the texture changes in fact result from a combination of rotation and some degree of pure shear strain.

To summarize, we have shown using molecular dynamics simulations that single-crystal tantalum shock-compressed along its $[101]$ direction undergoes something very close to rotation about $[10\bar{1}]$ thanks to a combination of preferential activation of one of two primary plasticity mechanisms of the types $[111](1\bar{2}1)$ or $[1\bar{1}1](121)$ and near-perfect cancellation of rotations arising from secondary plasticity mechanisms with Burgers vectors $[11\bar{1}]$ and $[\bar{1}11]$. This represents the first detailed account of how $[101]$ Ta might relax to state of low shear stress in the wake of shock while undergoing the lattice rotation observed in the dynamic compression experiment of Wehrenberg \emph{et.\ al.}\ \cite{Wehrenberg2017}. We have also shown that in isolated regions of the crystal, it is possible for the two primary mechanisms to become active in equal measure, leading to local stabilization of the compression axis. This physics will resurface in the context of copper shock-compressed along $[001]$ and $[111]$, which we now go on to discuss.

\subsection{\label{sec:001Cu} [001] Copper}

The next case we will consider is face-centered cubic (fcc) copper shocked along the $[001]$ direction ($[001]$ Cu). Like tantalum, copper is a prime testbed for plasticity studies thanks to the stability of its ambient phase at high pressures: the fcc phase persists up to at least 1~TPa under both static \cite{Greeff2006} and ramp-loading \cite{Fratanduono2020} conditions, and only transforms to a bcc structure upon shock compression at 180~GPa \cite{Sharma2020b}, shortly before it shock-melts at 230~GPa \cite{Hayes1999}. Below 180~GPa, copper relieves shock-induced shear stresses by a direction- and pressure-dependent combination of slip, twinning, and stacking-fault formation. According to MD simulations \cite{Holian1998,Germann2000,Germann2004,Cao2007} and shock-recovery experiments \cite{Meyers2003,Schneider2004,Cao2010}, these faults are particularly profuse when copper is loaded along $[001]$, and increase in density at greater pressures. Indeed, a recent study of shock-loaded polycrystalline copper performed by Sharma \emph{et.\ al.}\ \cite{Sharma2020b} exploited \emph{in situ} measurements of shifts in the $(200)$ diffraction peak to measure the stacking fault density as a function of pressure, using the classic formalism owed to Warren\cite{Warren1990} and Velterop \cite{Velterop2000}; they concluded that the typical fault density (averaged over the polycrystalline aggregate) increased steadily with compression until peaking at 10\% just below the fcc-bcc transition pressure. This study serves as an excellent demonstration of copper's potential as a model for fcc plasticity during dynamic compression.

One of the few studies to address the texture evolution of copper under shock is that of Suggit \emph{et.\ al.}\ \cite{Suggit2012}. The authors dynamically loaded copper single crystals along their $[001]$ axis up to pressures of 50~GPa and simultaneously used white-light (i.e.\ broadband) x-ray radiation to image their reciprocal-space structure. They observed that upon compression, the majority of the scattering peaks developed four diffuse `satellites' spaced regularly around them on the image plate. It was concluded that the $[001]$ axis was unstable to shock compression, and that each satellite was brought about by a different sub-domain of the crystal that had rotated either positively or negatively about its local $[100]$ or $[010]$ direction. Rotation of this kind is consistent with \emph{conjugate} (or \emph{duplex}) slip, in which pairs of $\{111\}$ slip planes (which, individually, cause rotation about $\langle110\rangle$-type axes) become active in equal measure, such that their rotations combine to give a net rotation about the $\langle100\rangle$ directions \cite{Hosford1960}. This work constituted the first direct evidence for slip-induced lattice rotations occurring in a shock-compressed single crystal on nanosecond timescales.

However, while the study of Suggit of coworkers succeeded in identifying the active slip planes in $[001]$ Cu, it did not attempt to reconcile the rotations observed with a physical model of plasticity-induced rotation. The authors did use a rotation model owed to Taylor \cite{Taylor1926,Taylor1927} (which, as discussed in Ref.~\cite{Heighway2021}, is technically unsuited to a planar shock-loading geometry, much like the Schmid model) to infer the post-shock dislocation density $\rho_d\ (\sim10^{12}\,\text{cm}^{-2})$, but did not use it to predict quantitatively the rotation angle expected along the Hugoniot for conjugate slip. It is therefore not yet established, to the knowledge of the authors, whether the conjugate slip systems detected by Suggit \emph{et.\ al.} are the \emph{sole} plasticity agents under shock, or whether, as in computational study of Avraam \emph{et.\ al.}\ \cite{Avraam2021}, there are additional active slip systems with which the two dominant systems are competing. In this section, we address this question by studying the defects nucleated under shock in $[001]$ Cu as predicted by large-scale MD simulations, and by interpreting these results and those of the above-mentioned experiment with a simple rotation model.

\begin{figure}[t]
    \includegraphics{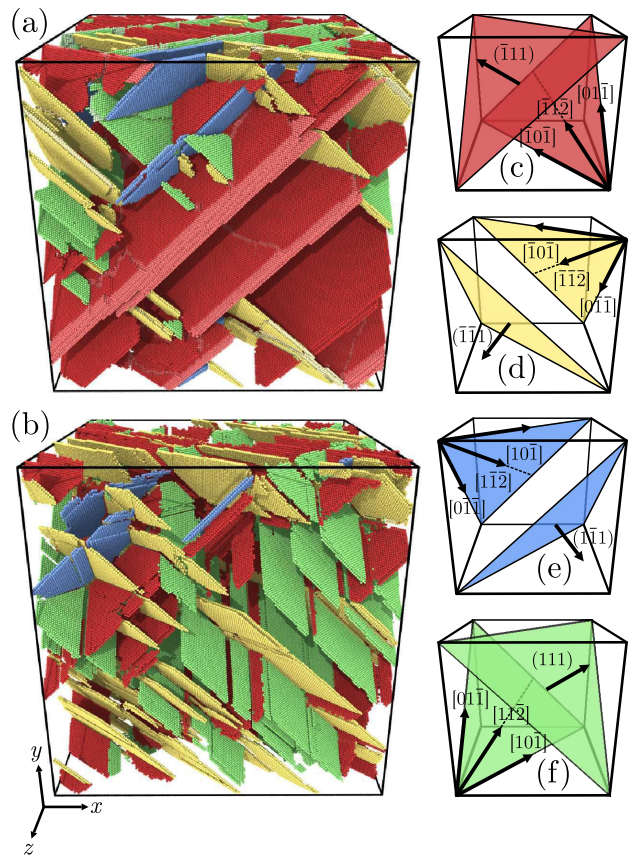}
    \caption{Plasticity mechanisms active in fcc copper shock-compressed along $[001]$ to 30~GPa. (a) Visualization of atoms situated on stacking faults or in deformation twins according to the template matching technique (TMT); atoms colored red, yellow, blue, and green are situated in faults or twins with habit planes $(\bar{1}11)$, $(\bar{1}\bar{1}1)$, $(1\bar{1}1)$, and $(111)$, respectively. Twinned material is colored slightly lighter for clarity. (b) Atoms situated on slip planes according to slip vector analysis (SVA), similarly colored according to their plane. (c-f) Simplified depiction of the fault, twin, and slip systems, of which 12 exist in total. These systems are enumerated in Tab.~\ref{tab:001CuSystems}.}
    \label{fig:001CuFaults}
\end{figure}

In Fig.~\ref{fig:001CuFaults}, we depict the range of plasticity mechanisms active in simulated $[001]$ Cu with its $\langle100\rangle$ directions initially aligned with the coordinate axes shock-compressed to 30~GPa. As expected of an fcc crystal, we find that all dislocation motion is restricted to the four close-packed $\{111\}$ planes, which are distributed symmetrically around the compression axis. For each of these planes, there exist three different directions in which shear motion takes place, two of the type $\{011\}$, and one of the type $\{112\}$. On the $(111)$ plane, for example, we observe a mixture of partial slip through vector $\frac{1}{6}[11\bar{2}]$ [giving rise to the stacking faults and deformation twins pictured in Fig.~\hyperref[fig:001CuFaults]{8(a)}] and full dislocation slip through vectors $\frac{1}{2}[10\bar{1}]$ or $\frac{1}{2}[01\bar{1}]$ [which brings about the slip planes shown in Fig.~\hyperref[fig:001CuFaults]{8(b)}]. There thus exist $4\times3 = 12$ active plasticity mechanisms in total, four of which are stacking fault and twin systems (which we will collectively call \emph{partial slip systems}), eight of which are full slip systems. To enumerate these systems, we adapt the notation used in Ref.~\cite{Shalaby1978}, in which the four distinct $\{111\}$ slip planes are labeled by letters a, b, c, and d, the three $\{101\}$ full slip directions by indices 1, 2, and 3. We denote the partial slip vector between vectors $i$ and $j$ by the index $ij$. Our notation is given in full in Tab.~\ref{tab:001CuSystems}.

\begin{table}[t]
\begin{tabularx}{8.5cm}{@{}YYYY@{}}
\hline
\hline
Plane $\mathbf{n}$ & Direction $\mathbf{m}$ & Mechanism(s) & Notation\\
\hline
\multirow{3}{*}[-0.5ex]{$(111)$} & $[01\bar{1}]$ & Slip & a$\bar{1}$\\
                         & $[10\bar{1}]$ & Slip & a$2$\\
                         & $[11\bar{2}]$ & Faults, twins & a$\bar{1}2$\\
\hline
\multirow{3}{*}[-0.5ex]{$(\bar{1}\bar{1}1)$} & $[0\bar{1}\bar{1}]$ & Slip & b$\bar{1}$\\
                                     & $[\bar{1}0\bar{1}]$ & Slip & b$2$\\
                                     & $[\bar{1}\bar{1}\bar{2}]$ & Faults, twins & b$\bar{1}2$\\
\hline
\multirow{3}{*}[-0.5ex]{$(\bar{1}11)$} & $[01\bar{1}]$ & Slip & c$\bar{1}$\\
                                       & $[\bar{1}0\bar{1}]$ & Slip & c$2$\\
                                       & $[\bar{1}1\bar{2}]$ & Faults, twins & c$\bar{1}2$\\
\hline
\multirow{3}{*}[-0.5ex]{$(1\bar{1}1)$} & $[0\bar{1}\bar{1}]$ & Slip & d$\bar{1}$\\
                                       & $[10\bar{1}]$ & Slip & d$2$\\
                                       & $[1\bar{1}\bar{2}]$ & Faults, twins & d$\bar{1}2$\\
\hline
\hline
\end{tabularx}
\caption{\label{tab:001CuSystems}Enumeration of the plasticity mechanisms active in shock-compressed $[001]$ Cu.}
\end{table}

The sense of the rotation brought about by these 12 plasticity mechanisms can be predicted using the same kinematic framework applied to $[101]$ Ta in Sec.~\ref{sec:101Ta}. We should emphasize that according to Eq.~(\ref{eq:NtoNalt}), the slip direction $\mathbf{m}$ only controls the \emph{magnitude} of the rotation suffered by any given crystal plane normal; it is the active slip plane normal $\mathbf{n}$ that dictates the \emph{direction} of the reorientation. That is to say that plasticity mechanisms a$\bar{1}$, a$2$, and a$\bar{1}2$ deflect reciprocal lattice vectors in exactly the same direction (when they cause deflection at all). There are therefore only as many distinct `rotation axes' as there are active slip planes, of which there are just four for fcc copper. This is in marked contrast to bcc tantalum studied in Sec.~\ref{sec:101Ta}, for which there were only four distinct slip \emph{directions}, but whose slip \emph{planes} were highly diverse.

\begin{figure}[t]
    \includegraphics{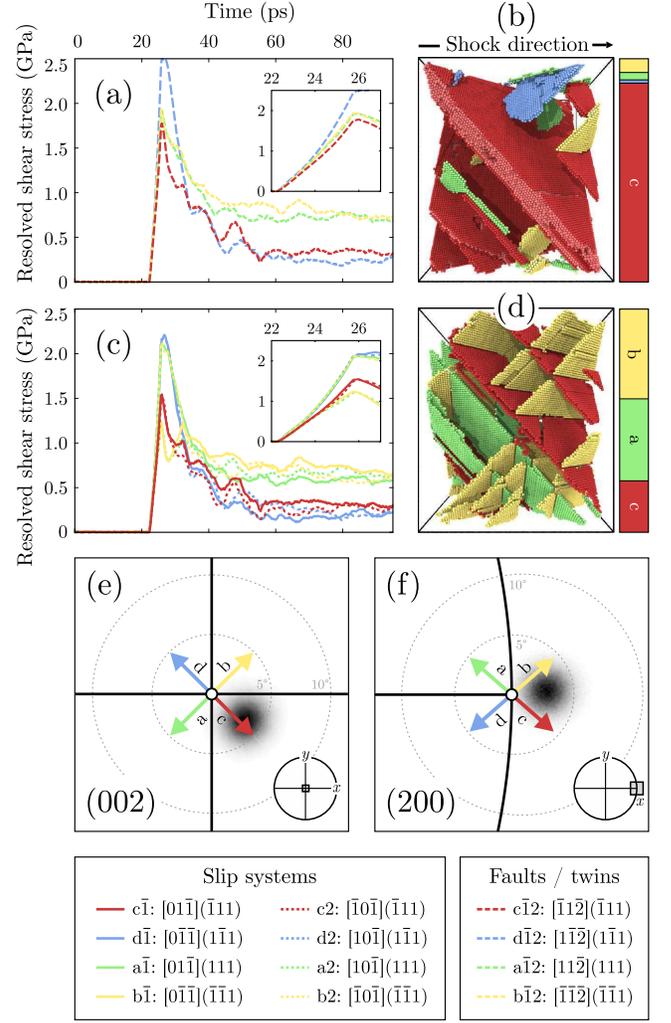}
    \caption{Behavior of a representative material element in $[001]$ Cu shock-compressed to 30~GPa. (a) Resolved shear stresses acting on twins and faults systems as functions of time. (b) Visualization of twins and faults at late times. (c,d) Resolved shear stresses and visualizations for the slip systems. The proportion in which the plasticity mechanisms are active is indicated by the stacked bar charts. (e,f) Pole plots showing late-time orientations of the $(002)$ and $(200)$ planes, respectively. White circles mark the initial orientation of these planes. Arrows indicate the sense in which the orientations should change if given plasticity mechanisms operate.}
    \label{fig:001CuLagrangian}
\end{figure}

To elucidate the net lattice rotation brought about by these plasticity mechanisms, we show in Fig.~\ref{fig:001CuLagrangian} a composite image analogous to Fig.~\ref{fig:101TaLagrangian} for a representative $[001]$ Cu material element compressed to 30~GPa and initially situated 125~nm from the piston. We show once more the resolved shear stresses driving the plasticity mechanisms as functions of time -- this time considering the full and partial slip systems separately -- alongside pole plots showing the direction of the forward $(002)$ planes and transverse $(200)$ planes. This figure reveals that the sense of the shock-induced rotation can be understood with reference to the individual activities of both the full and partial slip systems, which (in this instance) do not cause rotation in the same directions.

\begin{figure*}[t]
    \includegraphics{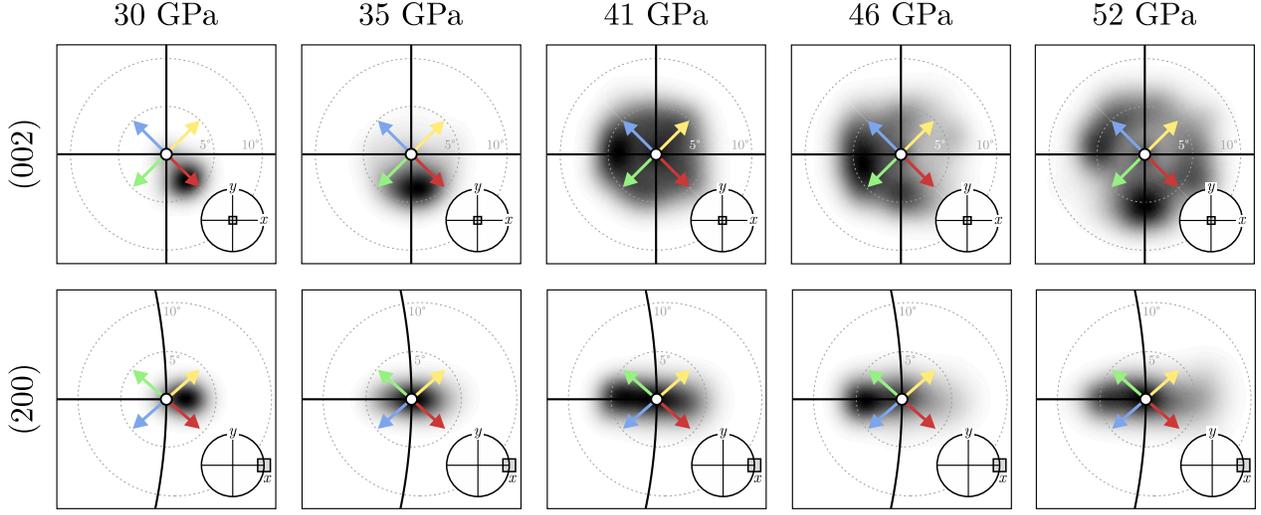}
    \caption{Pole plots showing the directions of the $(002)$ and $(200)$ planes in $[001]$ Cu shock-compressed to between 30 and 52~GPa shortly before shock breakout. Only the piston and a sliver of uncompressed material between the shock front and the rear surface have been excluded from the calculation of these pole plots. Arrows indicate the sense in which the orientations should change if given plasticity mechanisms operate.}
    \label{fig:001CuPressureScan}
\end{figure*}

\begin{figure}[b]
    \includegraphics{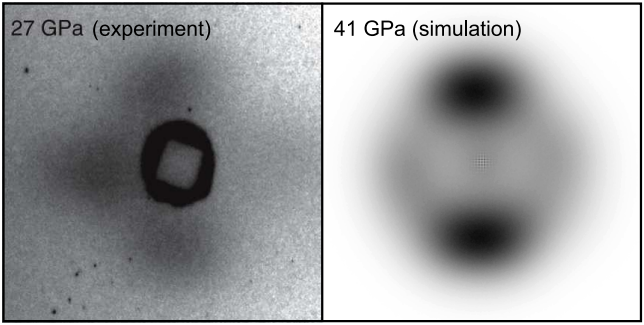}
    \caption{(Left) Image plate showing diffuse x-ray scattering features surrounding the (partially blocked) (002) peak of single-crystal crystal shocked along $[001]$ to 27~GPa from the experiment of Suggit \emph{et.\ al.}\ \cite{Suggit2012}. (Right) Pole plots showing the direction of the (002) planes in a simulated crystal shocked to 41~GPa, showing analogous diffuse satellite features. Left panel adapted with permission from Suggit \emph{et.\ al.}, Nature Communications \textbf{3}, 1224 (2012) \cite{Suggit2012}; copyright 2012 Nature Springer.}
    \label{fig:001CuComparison}
\end{figure}

The shock wave impacts the material element at $t = 22$~ps, consistent with the shock speed of $U_S\sim 5$~km\,s\textsuperscript{$-1$} expected of the 30~GPa Hugoniot state \cite{Bringa2004}. Shear stress accrues at an identical rate on each partial slip system and on each full slip system, as required by the crystal's initial fourfold rotational symmetry around the compression direction. Before the shock wave can completely traverse the element, this symmetry is broken by the onset of plasticity, which is mediated by a complex combination of partial slip and -- a few picoseconds later -- full slip. The partial slip response is dominated by mechanism c$\bar{1}2$ [which generates the thick deformation twin visible in Fig.~\hyperref[fig:001CuLagrangian]{9(b)}], while full slip is distributed more evenly between the systems with slip planes a, b, and c. Once plastic flow ceases and the shear stresses are relaxed as far as possible, the plasticity mechanisms fall into two sets: the first includes those with planes of the types a and b, which settle to a higher limiting resolved shear stress of approximately 0.7~GPa; the second comprises those with slip planes c and d, which experience a relatively low shear stress of 0.3~GPa or so. It is interesting to note that plasticity mechanisms d$\bar{1}$, d$2$, and d$\bar{1}2$ fall into this latter group despite their being almost completely inactive - the shear stress driving them is relieved entirely by the other plasticity mechanisms.

In Figs.~\hyperref[fig:001CuLagrangian]{9(e,f)}, we indicate the directions in which the $(002)$ and $(200)$ plane normals should rotate under the action of the plasticity mechanisms enumerated in Tab.~\ref{tab:001CuSystems}. As with tantalum, the proportion in which the plasticity mechanisms are activated can be used to explain (qualitatively, at least) the reorientation of these two crystal planes. Considering first the direction of $(002)$, we note that equal activation of mechanisms with slip planes a and b causes no change to its direction; the remaining activity on systems with slip plane c would then push the $(002)$ plane normal towards the $[1\bar{1}0]$ direction, which is exactly what one observes in the pole plot. By contrast, the equal activity on slip planes a and b pushes the $(200)$ plane normal `upwards' along the great circle joining the $x$- and $y$-directions (i.e., towards $[010]$); the remaining activity on slip plane c then drives the $(200)$ plane normal towards $[0\bar{1}\bar{1}]$, causing the net displacement towards the $[00\bar{1}]$ direction shown in the pole plot. The reorientation of the crystal bulk can thus be explained by the combination of active plasticity mechanisms.

As we move to higher shock pressures, we observe a qualitative change in the response of our simulated crystals. The pressure scan shown in Fig.~\ref{fig:001CuPressureScan} reveals that at 35~GPa, the $(002)$ plane rotates not towards $[1\bar{1}0]$, but towards $[0\bar{1}0]$. Meanwhile, the $(200)$ planes remain stationary, which suggests that the crystal is undergoing something close to rotation about its $[100]$ axis. At higher pressures yet, we notice that the distribution of $(002)$ plane normals becomes diffuse, indicating that the activity of the twelve plasticity mechanisms has become highly heterogeneous. However, close inspection of the pole plots reveals that while the orientation distribution is clearly delocalized, intensity tends to collect towards the $\pm[100]$ and $\pm[010]$ directions. In other words, different domains of the crystal locally rotate about either their $[100]$ or $[010]$ crystallographic axes, precisely the behavior observed by Suggit \emph{et.\ al.}\ \cite{Suggit2012}. The kinematics suggest that rotation of this kind is brought about by preferential activation of \emph{pairs} of slip planes, a deformation mode previously referred to as conjugate slip. Similar behavior was observed in simulated nanocrystalline Cu in the recent MD study by Hu \emph{et.\ al.}\ \cite{Hu2020} (though no connection with Suggit \emph{et.\ al.}'s experiment was made).

To highlight the similarity between the simulated and true behavior of $[001]$ Cu, we show in Fig.~\ref{fig:001CuComparison} a comparison between the $(002)$ pole plot from a considerably larger simulated crystal (with a $250\times250$ unit-cell cross-section) at 41~GPa and an image of the $(002)$ diffraction peaks for a copper crystal shocked to 27~GPa from the experiment of Suggit \emph{et.\ al.} The four satellite x-ray features from different rotated subdomains correspond directly to the maxima occurring in the synthetic pole plot. This simulated crystal happens to rotate predominantly about its $\pm[100]$ axes and thus accounts for two of the four experimental satellite peaks; repeated simulations with different thermal seeds can also bring about rotation around the $\pm[010]$ axes.

\begin{figure*}
    \includegraphics{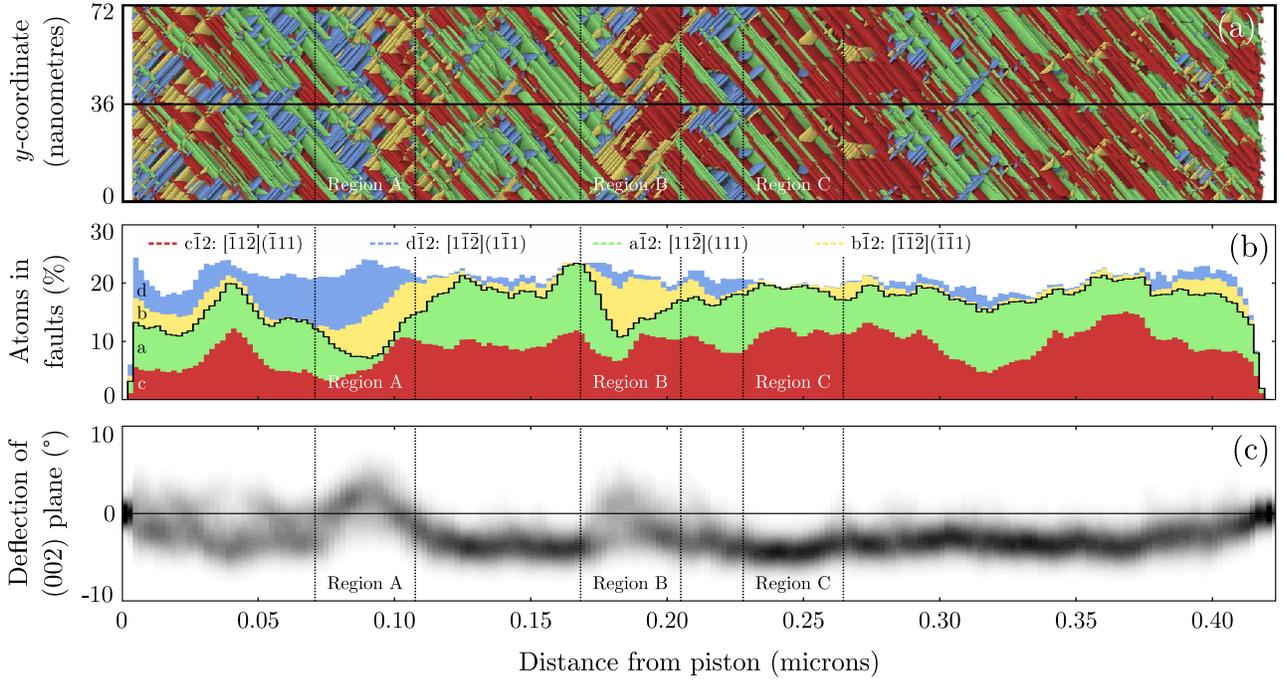}
    \caption{(a) Visualization of stacking faults a$\bar{1}2$, b$\bar{1}2$, c$\bar{1}2$, and d$\bar{1}2$ in $[001]$ Cu shocked to 35~GPa, shown along the length of the crystal. The piston-driven shock wave travels from left to right. One periodic image of the crystal is shown in the $y$-direction for clarity. (b) Fraction of atoms in each spatial bin belonging to the four stacking faults, displayed as a stacked bar chart. The black line separates contributions from $a\bar{1}2$ and $c\bar{1}2$ (which, together, cause rotation of $(002)$ towards $[0\bar{1}0]$) from contributions from $b\bar{1}2$ and $d\bar{1}2$ (which cause rotation towards $[010]$). (c) Distribution of deflection angles of the $(002)$ plane about the $x$-axis as a function of distance along the crystal. Regions B and C suffer close to minimal and maximal deflection, respectively, for this given shock pressure, while Region A shows `retrograde' rotation due to local dominance of $b\bar{1}2$ and $d\bar{1}2$.}
    \label{fig:001CuProfile}
\end{figure*}

To gain an idea of how `pure' the conjugate slip mode is, we now examine the variation of the local slip activities and attendant rotation along the length of the crystal at a shock pressure of 35~GPa. We show in Fig.~\hyperref[fig:001CuProfile]{12(a)} a visualization of the four stacking fault variants present shortly before breakout, and in Fig.~\hyperref[fig:001CuProfile]{12(b)} below it, we plot the faulted fraction of material as a function of $z$. We note that although we should of course also consider the activity of the full slip systems, their activity is strongly correlated with their corresponding partial slip system, meaning it suffices only to plot stacking faults for the purposes of understanding the net rotation. We observe that while slip planes a and c tend to dominate overall, there is throughout the crystal nonzero activity on the b and d planes. Wherever these `contaminant' slip planes are active, the prevailing rotation towards $[0\bar{1}0]$ is suppressed, as illustrated in Fig.~\hyperref[fig:001CuProfile]{12(c)}. In fact, competition between the two pairs of stacking faults leads to near-cancellation of the rotation in Region B, and in Region A, we observe localized \emph{reversal} of the net rotation direction due to plastic activity on planes b and d outstripping that on planes a and c. Our simulations therefore suggest that single-crystal copper shocked along $[001]$ will undergo imperfect conjugate slip, in which slip in concentrated largely -- but not exclusively -- onto two planes.

To better assess whether or not this imperfect conjugate slip mode is borne out by the experiment of Suggit \emph{et.\ al.}\ \cite{Suggit2012}, it is informative to study the simulated response of a $[001]$ Cu crystal that does undergo near-perfect conjugate slip. It  transpires that the crystal can be encouraged to do so simply by tilting its $[001]$ axis by as little as $1^\circ$ away from the compression axis before shocking it; doing so perturbs the initial shear stresses driving the four $\{111\}$ slip planes to such a degree that two are rendered almost completely inactive, thus bringing about near-ideal conjugate slip conditions. To achieve this pre-tilt while maintaining periodicity in the transverse directions, the crystal is oriented such that its $[100]$, $[0\,56\,\bar{1}]$, and $[0\,1\,56]$ directions are aligned with the $x$, $y$, and $z$ axes, respectively; the simulations parameters are in all other respects identical for these off-kilter crystals, whose behavior we now go on to explore.

\begin{figure}[t]
    \includegraphics{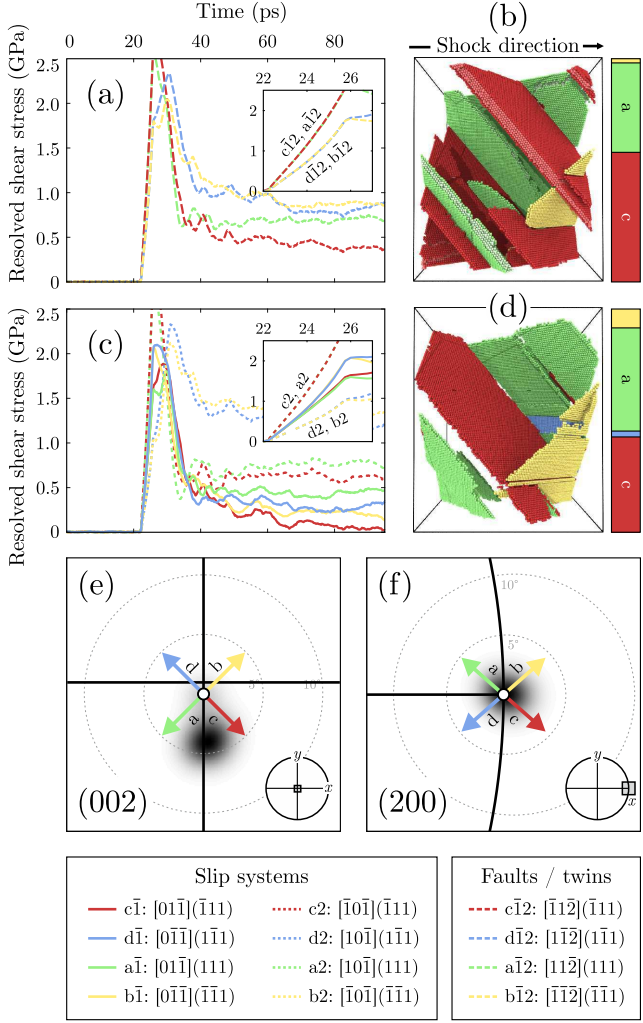}
    \caption{Behavior of a representative material element in $[001]$ Cu shock-compressed to 30~GPa and pretilted towards $[0\bar{1}0]$ by $1^\circ$. (a) Resolved shear stresses acting on twins and faults systems as functions of time. (b) Visualization of twins and faults at late times. (c,d) Resolved shear stresses and visualizations for the slip systems. The proportion in which the plasticity mechanisms are active is indicated by the stacked bar charts. (e,f) Pole plots showing late-time orientations of the $(002)$ and $(200)$ planes, respectively. White circles mark the initial orientation of these planes. Arrows indicate the sense in which the orientations should change if given plasticity mechanisms operate.}
    \label{fig:001CuLagrangianTilted}
\end{figure}

Fig.~\ref{fig:001CuLagrangianTilted} illustrates the evolution of a representative material element much like that considered in Fig.~\ref{fig:001CuLagrangian}, this time within a pretilted $[001]$ Cu crystal shocked to 30~GPa. We immediately see from the insets of Figs.~\hyperref[fig:001CuLagrangianTilted]{13(a,c)} that, in contrast to those in the perfectly-aligned crystal, the resolved shear stresses in this misaligned crystal do not increase at the same rates during the early stages of compression. More specifically, we observe that, before yield, the shear stresses rise most rapidly on partial slip systems a$\bar{1}2$ and c$\bar{1}2$ and on full slip systems a$2$ and c$2$. These elevated shear stresses cause almost exclusive activation of slip planes a and c [as shown in Figs.~\hyperref[fig:001CuLagrangianTilted]{13(b,d)}], which, in turn, causes the forward $(002)$ plane to rotate directly towards the $[0\bar{1}0]$ direction [as shown in the pole plots in Figs.~\hyperref[fig:001CuLagrangianTilted]{13(e,f)}].

\begin{figure*}[t]
    \includegraphics{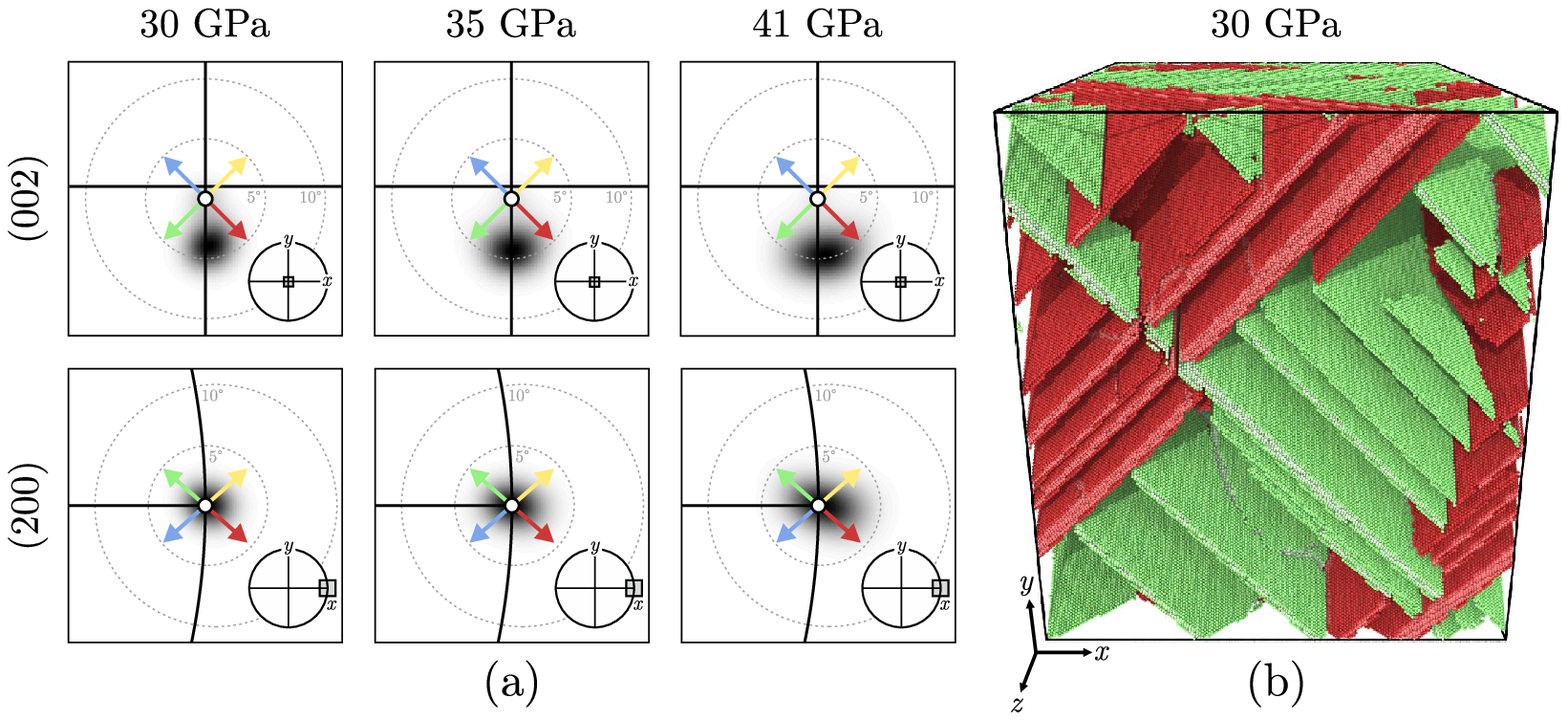}
    \caption{(a) Pole plots showing the directions of the $(002)$ and $(200)$ planes in $[001]$ Cu shock-compressed to between 30 and 41~GPa and pretilted towards $[0\bar{1}0]$ by $1^\circ$. Arrows indicate the sense in which the orientations should change if given plasticity mechanisms operate. (b) Visualization of stacking faults and deformation twins with habit planes a and c active at 30~GPa.}
    \label{fig:001CuPressureScanTilted}
\end{figure*}

The response of this misaligned crystal thus explains, in a simple sense at least, why $[001]$ Cu should be unstable to conjugate slip under compression. If the crystal is perturbed such that its $[001]$ direction rotates towards (say) $[0\bar{1}0]$, the attendant changes in shear stress are such as to favor slip on planes a and c, which precipitates further rotation towards the $[0\bar{1}0]$ direction, and so on. We also know that a uniaxially strained single crystal must deform on at least two independent slip systems to relax to a fully nondeviatoric elastic strain state. Therefore when a pair of slip planes inevitably becomes locally active in some region of the $[001]$ Cu crystal, it is apparently more likely, on the basis of the resolved shear stresses alone, for plastic flow to continue on that pair of planes than for the previously inactive slip systems to become operative. For the pretilted crystal, this positive feedback leads to near-perfect conjugate slip that is reproducible over a range of pressures, as illustrated in the pole plots in Fig.~\ref{fig:001CuPressureScanTilted}. For the perfectly-aligned crystal, the feedback effect cannot eliminate the contaminant slip planes entirely, but it is still strong enough that regions of conjugate slip can be clearly discerned in the pole plots pictured in Fig.~\ref{fig:001CuPressureScan}.

The results of these simulations naturally prompt the question: with what level of contamination are the lattice rotations measured by Suggit \emph{et.\ al.}\ consistent? That the copper crystals in their experiment underwent conjugate slip following shock-compression is evident; what is less clear is whether the measured amount of rotation is consistent with \emph{pure} conjugate slip.

\begin{figure}[b]
    \includegraphics{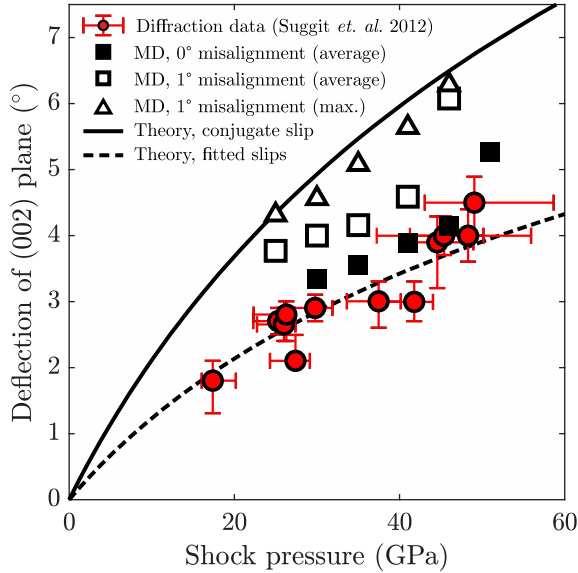}
    \caption{Deflection of the $(002)$ planes around the $\langle100\rangle$ directions against shock pressure in $[001]$ Cu. Data from Suggit \emph{et.\ al.}\ (circles, Ref.~\cite{Suggit2012}) are compared with simulations of crystals shocked directly along $[001]$ (solid squares) and along a direction at $1^\circ$ to $[001]$ (open squares and triangles). Lines show theoretical deflection based on minimization of the deviatoric elastic strain for perfect and imperfect conjugate slip (the latter with impurity $\alpha = 0.28$) (solid and dashed lines, respectively).}
    \label{fig:001CuRotation}
\end{figure}

To address this question, we compare in Fig.~\ref{fig:001CuRotation} the experimental measurements of shock-induced rotation obtained by Suggit \emph{et.\ al.}\ with the results of our MD simulations, considering both the perfectly-aligned and misaligned crystals. For the perfectly-aligned $[001]$ Cu, meaningful comparisons with the data are only possible above a cut-off pressure of 35~GPa, below which the crystal did not reliably undergo conjugate slip (as was shown in Fig.~\ref{fig:001CuPressureScan}). The typical rotations we extract from these ideal crystals (which we do by locating the maxima of the pole plot intensity distribution) increase monotonically with shock pressure, and compare reasonably well with the experimental data. By contrast, we observe that the deflection suffered by the $(002)$ plane for the crystals tilted by $1^\circ$ before compression is systematically higher than that of the perfectly-aligned crystals, and is in fact so high as to be inconsistent with the data to within experimental error. The salient difference between the two sets of simulations is of course the purity of the conjugate slip. We suggest that the experimental data are better explained by the simulations of perfectly-aligned $[001]$ Cu crystals, for which the conjugate slip is relatively impure.

To add weight to this argument, we can use the kinematic equations defined in Sec.~\ref{sec:101Ta} [Eqs.~(\ref{eq:elastoplastic},\ref{eq:plasticdef},\ref{eq:totaldef})] to predict the amount of deflection one would expect of a $[001]$ Cu crystal if it were to undergo perfect conjugate glide. To do so, we choose a pair of non-opposing $\{111\}$ slip planes, and assume that the mean slip direction on each is simply the unweighted average of the three active slip directions $\bar{1}$, $2$, and $\bar{1}2$ (which is $\bar{1}2$). Then, for a given imposed compression ratio $V/V_0$, we calculate the locus of accessible elastic deformation states $F^e(\gamma|V)$ as a function of the glide $\gamma$ on two active slip systems. To calculate the slip activity $\gamma$ `properly' for a given amount of compression would of course require a high-pressure constitutive model of copper that accounts for the critical resolved shear stress of every slip system, the elastic constants, work-hardening rates, and so on. Rather than calculate $\gamma$ in this way, we opt instead to seek the value of $\gamma$ that minimizes the von Mises effective elastic strain $\varepsilon_{\text{vms}}^e$, defined as
\begin{equation}
    \label{eq:vms}
    \varepsilon_{\text{vms}}^e = \sqrt{\frac{2}{3}\,\varepsilon^e : \varepsilon^e},
\end{equation}
where $:$ is the double inner product operator, and $\varepsilon^e$ is the deviatoric part of the right elastic stretch tensor $U^e$, obtained from the right polar decomposition of $F^e$:
\begin{equation}
    \label{eq:decomp}
    F^e = RU^e.
\end{equation}
The effective strain $\varepsilon_{\text{vms}}^e$ is a scalar, frame-invariant measure of the distortion of the crystal's unit cell that we seek to minimize based on the notion that shock-compressed metals generally relax to a state of minimal deviatoric elastic strain. This principle does indeed appear to hold for the crystals simulated here: taking the pretilted crystal shocked to 30~GPa as an example, we find that following the initial uniaxial compression by factor $V/V_0 = 0.86$ -- which, in the absence of plasticity, amounts to an effective strain of $9.47\%$ -- the conjugately-slipped regions of the crystal relax to $\varepsilon_{\text{vms}}^e = 2.16\%$, where, on the basis of the kinematics, the minimal effective shear strain attainable should be $2.13\%$. Minimization of the effective strain therefore appears to be a reasonable basis for our semi-empirical rotation model.

Once calculated, the slip activity $\gamma$ that minimizes $\varepsilon_{\text{vms}}^e$ is then used to calculate $F^e$, which transforms crystal plane normals via $\mathbf{N} \to [(F^e)^T]^{-1} \mathbf{N}$. This approach to calculating the deflection as a function of compression allows us to formulate rotate predictions quickly without needing to appeal to a complex crystal plasticity model. We note in passing that this method of minimizing the deviatoric elastic strain is similar in spirit to the rotation model employed by Wehrenberg \emph{et.\ al.}\ \cite{Wehrenberg2017} (in which the diagonal elements of $F^e$ were forced to be equal), but is done here in such a way that internal kinematic consistency is guaranteed.

The theoretical deflection expected under perfect conjugate glide and minimal effective strain conditions is plotted as a function of pressure in Fig.~\ref{fig:001CuRotation}. We see that the deflection prediction does \emph{not} successfully reproduce the average rotation of even the pretilted crystals. However, we must remember that even in these misaligned crystals, the conjugate slip is not perfect everywhere; if instead we identify the isolated regions of the pretilted crystals in which the conjugate slip is of the highest purity, where the attendant rotation is maximal, we find that the rotation prediction assuming ideal conjugate glide is very successful. This agreement reinforces the idea that the simulated crystals at least can indeed be reasonably modeled as deforming such as to minimize their effective elastic strain $\varepsilon_{\text{vms}}^e$. If we assume that the real crystals studied by Suggit and coworkers \emph{also} deform in this manner, we are forced to conclude that they cannot be deforming by perfect conjugate slip.

To quantify the extent to which the conjugate slip is imperfect, we can adapt our simple rotation model by allowing nonzero, equal activity on the two contaminant slip systems. Specifically, we can say that if the activity on the two dominant slip systems is $\gamma$, the remaining two slip systems are active to the extent $\alpha\gamma$, where $\alpha$ parametrizes the degree of impurity; $\alpha=0$ denotes perfect conjugate glide (and therefore maximal rotation), while $\alpha=1$ means equal activation of all four slip planes (hence no rotation). If we assume for convenience that $\alpha$ does not change with shock pressure -- though there is no obvious reason to expect this will be the case -- and calculate the value of $\alpha$ that best explains the rotation data based on a maximum likelihood estimation, we obtain the second fit shown as a dashed curve in Fig.~\ref{fig:001CuRotation}. This fit accounts for the experimental dataset reasonably well over the whole pressure range, and is characterized by an impurity parameter of $\alpha = 0.28$. In other words, we suggest that the data of Suggit \emph{et.\ al.} are consistent with imperfect conjugate slip for which the two dominant slip planes accommodate approximately $1/(1+\alpha) \sim 80\%$ of the total plastic strain.

We should note that there is in principle a second way of explaining the `small' rotations observed in the experiment: it is conceivable that the conjugate glide \emph{is} near-perfect, but that the strength of the crystals is far greater than the simulations would suggest. Broadly speaking, a stronger crystal stops plastically deforming earlier along its deformation path, and therefore suffers less rotation than a weaker crystal whose path takes it nearer the hydrostat. However, calculations of the strength needed to bring about the observed level of rotation suppression seem to rule out this possibility. Taking a crystal shocked to 30~GPa as an example once again, we can calculate the conjugate slip activity $\gamma$ required to bring about the observed deflection of $\sim 3^\circ$ (which is only about 60\% of that required to minimize $\varepsilon_{\text{vms}}^e$ at this pressure). The elastic deformation gradient $F^e$ computed from this value of $\gamma$ yields longitudinal and transverse elastic strains of 8.3\% and 2.8\%, respectively. If we then construct an ideal crystal with this same strain state in an MD simulation and model its interactions under the Mishin potential, we find that the crude shear stress, conventionally defined as
\begin{equation}
    \label{eq:crude_shear}
    \tau = \frac{1}{2}\left[\sigma_{zz} - \frac{1}{2}(\sigma_{xx} + \sigma_{yy})\right],
\end{equation}
takes a value of 1.9~GPa. By contrast, when Murphy \emph{et.\ al.}\ used x-ray diffraction to measure the shear strength of copper single crystals shocked along $[001]$ to the considerably higher pressure of $(112\pm5)$~GPa, they inferred a shear stress of just $(1.1\pm0.2)$~GPa \cite{Murphy2010}. Given that material strength is generally expected to increase as one moves up the Hugoniot \cite{Comley2013,Brown2014}, it seems highly unlikely a shear stress of 1.9~GPa could be supported at 30~GPa shock pressure. We therefore maintain that it is competition between different slip systems, rather than extreme strength, that is responsible for the relatively low rotation measured in Suggit \emph{et.\ al.}'s experiment \cite{Suggit2012}.

One final question we wish to address is: does the identity of the active slip \emph{directions} influence our conclusion? We assumed when constructing our rotation model that the average slip direction on each $\{111\}$ slip plane is of the type $\langle112\rangle$, for which the resolved shear stress is maximal. This assumption is reasonable for copper modeled under the Mishin potential, for which the stacking fault density is exceptionally high, reaching around 30\% at 50~GPa shock pressure. However, the recent study of Sharma \emph{et.\ al.} suggests that the fault fraction at this pressure should be closer to 4\% \cite{Sharma2020}, which implies that a considerably greater fraction of the plastic strain is in reality mediated by full slip. One could then ask whether the dominant slip directions being of the type $\langle011\rangle$ rather than $\langle112\rangle$ substantively alters the implications of the model.

\begin{figure*}[t]
    \includegraphics{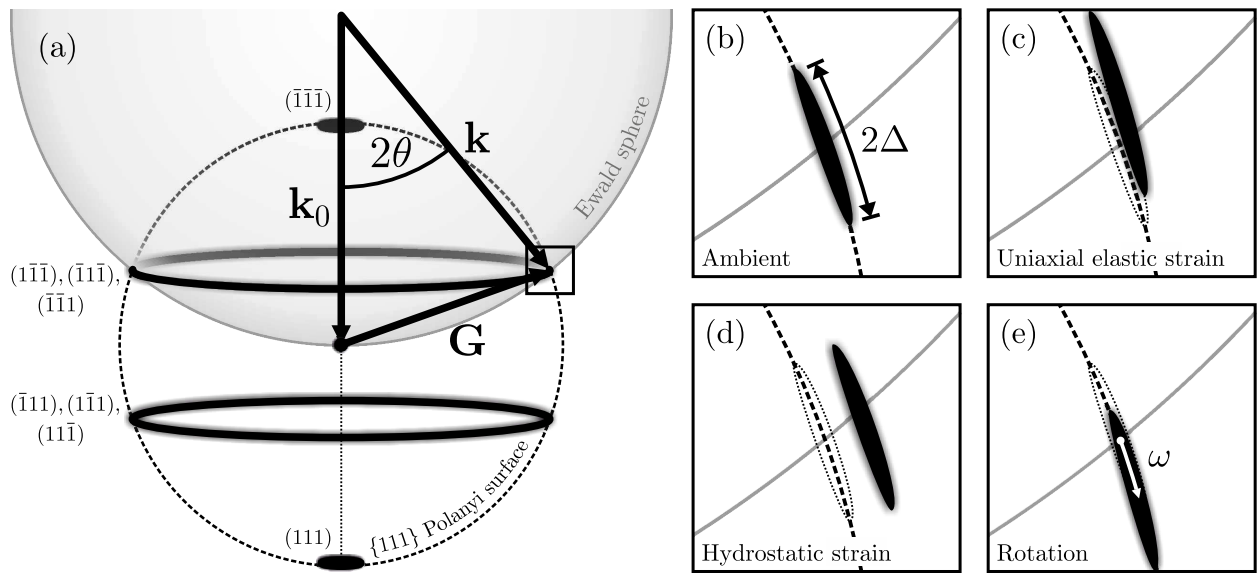}
    \caption{Reciprocal-space illustration of the scattering geometry for the experiment of Milathianaki \emph{et.\ al.}\ \cite{Milathianaki2013}, in which $[111]$-fiber-textured copper was probed by an 8.8~keV x-ray beam traveling parallel to the fiber axis. Shown to scale are the Ewald sphere and the $\{111\}$ Polanyi surface, which would be completely and uniformly occupied by $\{111\}$ scattering vectors in a powderlike sample; due to the sample's fiber texture, the $\{111\}$ scattering vectors are found only at the poles and on narrow rings situated at $\pm19.5^\circ$ latitude. (b-e) Close-ups of the intersection between the Ewald sphere and the actively scattering $\{111\}$ vectors under ambient conditions, following purely elastic strain along the fiber axis, in a state of purely hydrostatic elastic strain, and after pure rotation, respectively. Rotation through an angle $\omega$ comparable to or exceeding the texture half-width $\Delta$ moves the $\{111\}$ planes out of the Bragg condition, attenuating the diffraction pattern.}
    \label{fig:polanyi}
\end{figure*}

We note that the original model can still be applied unaltered if the two perfect dislocations on each operative slip plane were locally active to the same extent: the kinematic equations we have used to describe plasticity-induced rotation cannot `tell the difference' between shear in the $[11\bar{2}]$ direction mediated by (say) stacking fault a$\bar{1}2$ and shear mediated in the $[01\bar{1}] + [10\bar{1}] = [11\bar{2}]$ direction by equal amounts of activity on full slip systems a$\bar{1}$ and a$2$. In other words, the rotation depends only upon the net slip direction, not upon which combination of events leads to that slip. That said, it is entirely possible that the full slip directions are \emph{not} equally represented on each active slip plane, and that the sole active plasticity mechanisms are, for instance, a$2$ and c$\bar{1}$. When we re-apply our rotation model, this time assuming there is one active $\langle011\rangle$ dislocation on each of the two active slip planes, we find that the crystal cannot relax to so great an extent as before: the minimum achievable value of $\varepsilon_{\text{vms}}^e$ is about double that achievable for slip along $\{112\}$. However, the rotation curve is very similar, with the $(002)$ peak being deflected by only 10\% less than it would be for $\{112\}$-type slip. Even for this alternative form of conjugate slip, then, we would conclude on the basis of the small experimental rotations that the conjugate slip mode must be impure.

The level of impurity implied by our rotation model contains valuable information about the kinetics of plasticity in dynamically loaded copper. In the aforementioned study of $[011]$ Ta by Avraam \emph{et.\ al.}\ \cite{Avraam2021}, it was found that the contrast between the activities of the two primary plasticity mechanisms -- which controlled the net rotation -- was a highly non-linear function of the applied stresses governed by a range of kinetic effects. In particular, the effects of pressure hardening and homogeneous dislocation nucleation `rewarded' whichever of the two primary mechanisms was triggered first, leading to a feedback effect whereby the contrast became exacerbated at greater shock pressures. We suggest that a similar study of copper using the data of Suggit \emph{et.\ al.}, wherein the lattice rotation is used to constrain the slip activities and so the underlying dislocation kinetics, would be highly profitable.

In summary, we have conducted simulations of single-crystal copper shocked along $[001]$, and have observed preferential activation of pairs of slip planes of the type $\{111\}$ consistent with conjugate slip. The direction of the attendant rotation was found to agree with that observed in the dynamic-compression experiment of Suggit \emph{et.\ al.}\ \cite{Suggit2012}. By constructing a simple rotation model based on minimization of the post-shock deviatoric elastic strains, we were able to give a reasonable quantitative account of the experimental data by assuming that only 80\% of the plasticity is borne by the two dominant slip planes; the remaining 20\% is mediated by the two other slip planes, whose effect is to reduce the net rotation suffered by the crystal structure.

\subsection{\label{sec:111Cu} [111] Copper}

The third and final case we will study is copper shocked along $[111]$. This choice is motivated by the dynamic compression experiment of Milathianaki \emph{et.\ al.}\ \cite{Milathianaki2013}, in which copper polycrystals whose grains had their $[111]$ directions preferentially aligned with the sample normal were rapidly compressed to a peak normal stress of $73$~GPa with a $170$-ps-long short-pulse optical laser and probed with a femtosecond x-ray laser at 10-ps intervals. By analyzing the intensity profile of the $\{111\}$ Debye-Scherrer ring with the aid of a hydrocode, the authors could infer the stress and strain distributions in the copper foils as functions of time, allowing them to watch the crystal undergo one-dimensional elastic compression before relaxing plastically to a state of reduced deviatoric elastic strain over just a fraction of a nanosecond.

While texture evolution was not the main focus of the experiment, the fact that diffraction was visible at all places constraints on the amount of rotation caused by the compression process. When a fiber-textured polycrystal is illuminated by a quasi-monochromatic beam of x-rays, one generally observes not complete Debye-Scherrer rings but pairs of diffraction peaks distributed symmetrically around each ring. Reorientation of the crystal structure causes these peaks to migrate along the azimuthal direction (that is, along the ring) by an amount proportional to the lattice rotation. This was the means by which Wehrenberg \emph{et.\ al.}\ measured the plasticity-induced rotation as a function of shock pressure in tantalum nanocrystals \cite{Wehrenberg2017}. For this scheme to work, the samples must be oriented such that the incoming x-ray beam makes an angle with the fiber direction (this angle being $35^\circ$ in the above-named experiment).

If instead the x-ray beam is launched directly along the fiber axis -- as was done in the experiment of Milathianaki \emph{et.\ al.}\ -- the nature of the resulting diffraction pattern is materially different. A schematic depiction of the scattering geometry is given by Fig.~\ref{fig:polanyi}. In such a colinear geometry, one will generally see no diffraction at all unless the beam's wavelength is chosen precisely so that the Ewald sphere intersects a ring of scattering vectors. Tuning the beam energy this way allows one to generate one (and usually only one) complete Debye-Scherrer ring; in the present experiment, the authors tuned the XFEL photon energy to 8.8~keV to bring a subset of $\{111\}$ planes into the diffraction condition. During compression, changes to the sample's elastic strains cause the scattering vectors to migrate, allowing one to characterize the sample's response to rapid dynamic compression via the motion of its diffraction peaks. For instance, Milathianaki \emph{et.\ al.}\ were able to distinguish the uniaxial elastic strain state found in the elastic precursor wave [which, as shown in Fig.~\hyperref[fig:polanyi]{16(c)}, causes a marginal increase the the average scattering angle $2\theta$] from the more hydrostatic elastic strain state accompanying plastic relaxation [which causes a far greater shift in $2\theta$, as shown in Fig.~\hyperref[fig:polanyi]{16(d)}].

However, the changes to a crystal's elastic strain state that can be measured by this experimental arrangement are ultimately limited by the sample's finite texture width. Considering again Fig.~\hyperref[fig:polanyi]{16(c)}, we see that if the elastic strains within a precursor wave become too extreme, the entire distribution of $\{111\}$ scattering vectors will be pushed out of the Bragg condition, leading to complete extinction of the diffraction pattern. A similar problem is posed by plasticity-induced rotation: if, as in Fig.~\hyperref[fig:polanyi]{16(d)}, the scattering vectors turn through an angle $\omega$ that approaches the original texture half-width $\Delta$, the intensity of the reciprocal lattice along the Ewald sphere will diminish severely, practically extinguishing the Debye-Scherrer ring. However, the $\{1\bar{1}1\}$ ring studied by Milathianaki \emph{et.\ al.}~was in fact visible both before and during compression. The persistence of the diffraction thus hints at the possibility that the plasticity detected in these $[111]$ Cu targets was \emph{not} accompanied by appreciable rotation, at least over the timescales probed by the experiment. Our first objective is to establish whether this behavior is borne out by MD simulations, and, if so, to understand why.

\begin{figure}[b]
    \includegraphics{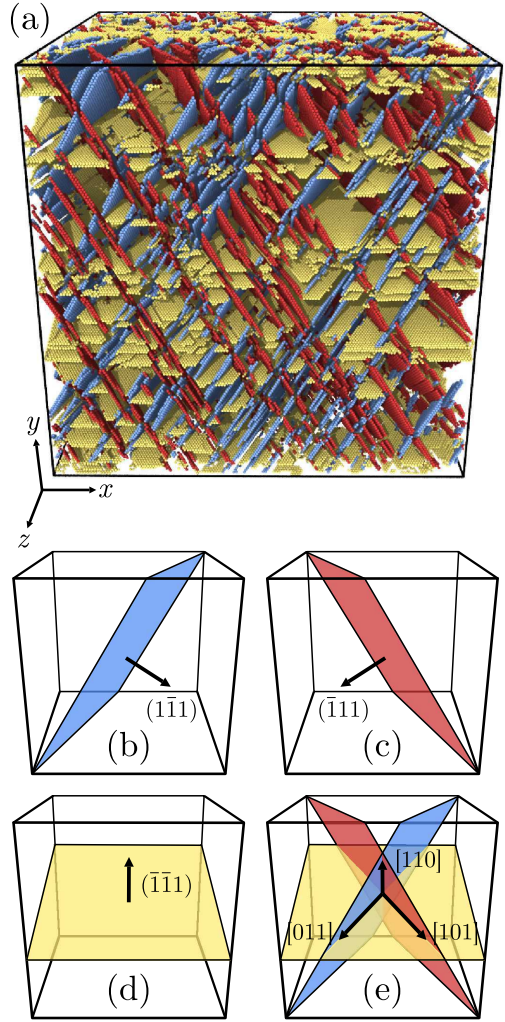}
    \caption{Plasticity mechanisms active in fcc copper shock-compressed along $[111]$ to 47~GPa. (a) Visualization of atoms situated on slip planes according to slip vector analysis (SVA). Atoms colored red, yellow, and blue are situated on planes $(\bar{1}11)$, $(\bar{1}\bar{1}1)$, and $(1\bar{1}1)$, respectively. (b-d) Simplified depiction of the slip planes. (e) Superposition of the active slip planes showing the three slip directions shared between them. These systems are enumerated in Tab.~\ref{tab:111CuSystems}}
    \label{fig:111CuSlips}
\end{figure}

\begin{table}[b]
\begin{tabularx}{8.5cm}{@{}YYYY@{}}
\hline
\hline
Plane $\mathbf{n}$ & Direction $\mathbf{m}$ & Mechanism(s) & Notation\\
\hline
\multirow{2}{*}[-0.5ex]{$(11\bar{1})$} & $[0\bar{1}\bar{1}]$ & Slip & b$1$\\
                                     & $[\bar{1}0\bar{1}]$ & Slip & b$\bar{2}$\\
\hline
\multirow{2}{*}[-0.5ex]{$(\bar{1}11)$} & $[\bar{1}0\bar{1}]$ & Slip & c$2$\\
                                       & $[\bar{1}\bar{1}0]$ & Slip & c$\bar{3}$\\
\hline
\multirow{2}{*}[-0.5ex]{$(1\bar{1}1)$} & $[\bar{1}\bar{1}0]$ & Slip & d$\bar{3}$\\
                                       & $[0\bar{1}\bar{1}]$ & Slip & d$\bar{1}$\\
\hline
\hline
\end{tabularx}
\caption{\label{tab:111CuSystems}Enumeration of the plasticity mechanisms active in shock-compressed $[111]$ Cu.}
\end{table}

We show in Fig.~\ref{fig:111CuSlips} the plasticity mechanisms detected in a simulation of a copper single crystal shock-compressed along $[111]$ to 47~GPa, whose $[1\bar{1}0]$, $[11\bar{2}]$, and $[111]$ directions were originally aligned with the $x$-, $y$-, and $z$-axes, respectively. Analysis of Fig.~\ref{fig:111CuSlips} immediately reveals that the plastic response of $[111]$ Cu differs from that of $[001]$ Cu in two regards. First, the plastic strain in $[111]$ Cu is realized almost exclusively by full dislocation slip. While stacking faults do form immediately behind the plastic front, the orientation of the crystal with respect to the loading axis is such as to favor the emission of a second, trailing partial dislocation on the same fault plane that `closes' the fault; detailed analysis of the formation of these stacking-fault ribbons in Mishin copper may be found in Ref.~\cite{Germann2004}. The second difference is the number of active slip systems. By aligning the $[111]$ direction with the compression axis, the (initial) shear stress acting on the $(111)$ plane is reduced to zero, rendering it inactive. Each of the remaining operative slip planes (b, c, and d) supports slip in two of three $\frac{1}{2}\langle110\rangle$ directions, which are `shared' among the slip planes -- the geometry is shown in Fig.~\hyperref[fig:111CuSlips]{17(e)}. The six active slip systems in $[111]$ Cu are tabulated in Tab.~\ref{tab:111CuSystems}.

\begin{figure}[b]
    \includegraphics{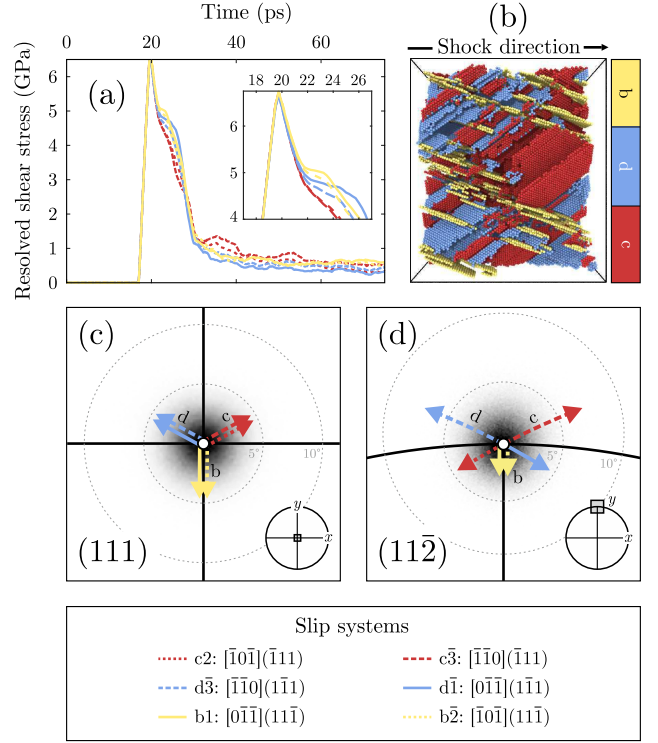}
    \caption{Behavior of a representative material element in $[111]$ Cu shock-compressed to 47~GPa. (a) Resolved shear stresses acting on the slip systems as functions of time. (b) Visualization of slip planes at late times. The proportion in which the plasticity mechanisms are active is indicated by the stacked bar chart. (c,d) Pole plots showing late-time orientations of the $(111)$ and $(11\bar{2})$ planes, respectively. White circles mark the initial orientation of these planes. Arrows indicate the sense in which the orientations should change if given plasticity mechanisms operate. The lengths of the arrows are proportional to the amount of reorientation resulting from a fixed amount of glide $\gamma$.}
    \label{fig:111CuLagrangian}
\end{figure}

\begin{figure*}[t]
    \includegraphics{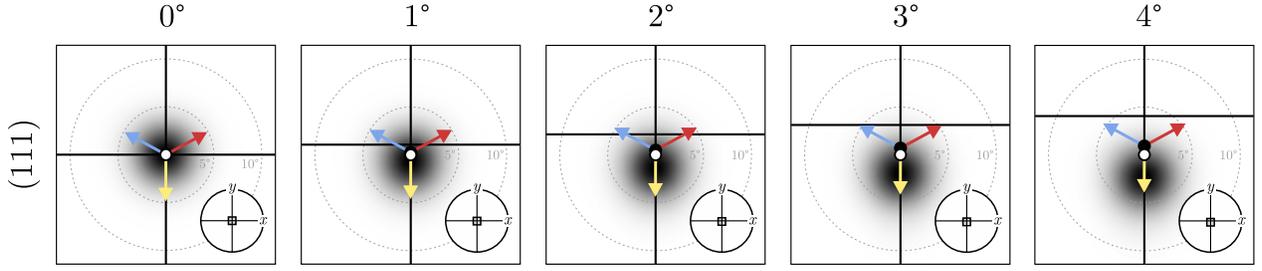}
    \caption{Pole plots showing the direction of the $(111)$ plane in $[111]$ Cu shock-compressed to 70~GPa and pretilted by angles between $0^\circ$ and $4^\circ$ around the $x$ axis, shortly before shock breakout. Only the piston and a sliver of uncompressed material between the shock front and the rear surface have been excluded from the calculation of these pole plots. White circles mark the initial orientation of the $(111)$ plane. Black circles mark its orientation after uniaxial compression, but before plasticity. Arrows indicate the sense in which the orientations should change if given plasticity mechanisms operate.}
    \label{fig:111CuDeflectionScan}
\end{figure*}

When we examine a representative material element in shocked $[111]$ Cu, we find that its rotation behavior differs starkly from that of both $[101]$ Ta and $[001]$ Cu. Fig.~\ref{fig:111CuLagrangian} shows the shear stress evolution, active slip planes, and rotation state of an element of copper initially situated 125~nm from the piston. When the shock wave passes the element at $t=18$~ps, shear stress increases at an equal rate on the six active slip systems, as shown in Fig.~\hyperref[fig:111CuLagrangian]{18(a)}, owing to the crystal's threefold rotational symmetry about the compression direction. When the crystal yields, the rate of shear-stress relief and the limiting value of the shear stress reached by each slip system is similar, implying the activity on each slip plane is almost identical. This is confirmed by the results of the slip-vector analysis pictured in Fig.~\hyperref[fig:111CuLagrangian]{18(b)}, which shows that the amounts of slip on planes b, c, and d are nearly equal. Unlike $[101]$ Ta and $[001]$ Cu, then, there is no preferential activation of certain slip systems at even the local level in $[111]$ Cu -- all three active slip planes pervade the entire crystal.

\begin{figure}[b]
    \includegraphics{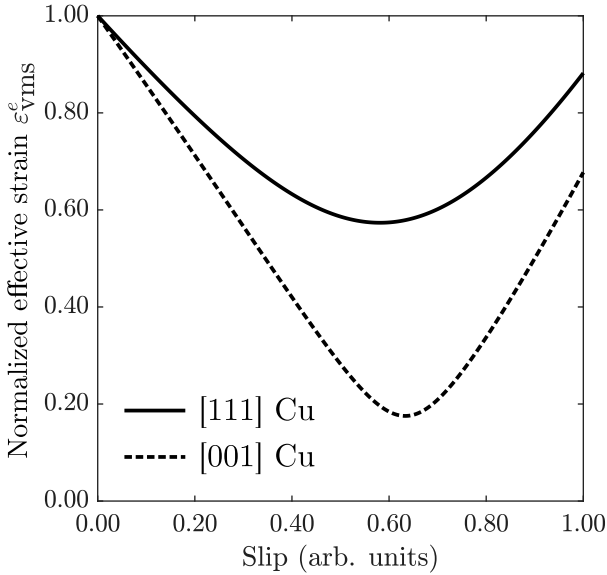}
    \caption{Variation of von Mises effective elastic strain $\varepsilon_{\text{vms}}^e$ with slip for copper crystals gliding on two $\{11\bar{1}\}$ planes following 25\% uniaxial compression. Slip activity is expressed in arbitrary units and $\varepsilon_{\text{vms}}^e$ is normalized to the value it assumes before slip takes place. The crystal compressed along $[111]$ is permitted to slip equally on systems c$2$, c$\bar{3}$, d3, and d$\bar{1}$ (see Tab.~\ref{tab:111CuSystems} for notation); the crystal compressed along $[001]$ is allowed to slip equally on systems c$\bar{1}$, c$2$, d$\bar{1}$, and d$2$ (see Tab.~\ref{tab:001CuSystems} for notation).}
    \label{fig:locicomparison}
\end{figure}

In Figs.~\hyperref[fig:111CuLagrangian]{18(c,d)}, we show pole plots for the forward $(111)$ and transverse $(11\bar{2})$ planes overlaid with arrows indicating the direction in which each slip system causes these planes to rotate. Focusing on the forward direction, we see that the three slip planes pull the $(111)$ plane normal in different directions related by $120^\circ$ rotations about $z$. If the three slip planes are active to the same extent, it would imply that, by symmetry, there should be no net deflection of the $(111)$ planes. This is exactly the behavior revealed by the pole plot. It turns out that while the geometry is slightly more complex, exactly the same reasoning holds for the transverse $(11\bar{2})$ planes, which are also undeflected. In fact, we can use Eq.~(\ref{eq:plasticdefpoly}) to show that if all six slip systems have identical activity $\gamma$, the resulting plastic deformation gradient takes the form
    \begin{equation}
        \label{eq:111hydro}
        F^p = \begin{pmatrix}1 + \frac{1}{3}(\sqrt{6}\gamma) & 0 & 0 \\ 0 & 1 + \frac{1}{3}(\sqrt{6}\gamma) & 0 \\ 0 & 0 & 1 - \frac{2}{3}(\sqrt{6}\gamma)\end{pmatrix}.
    \end{equation}
The equal activation of all six systems observed in the simulations thus produces a diagonal plastic strain state that requires no rotation of the crystal structure and leaves the direction of the atomic planes normal to the coordinate axes entirely undisturbed. We have repeated these simulations at higher piston (particle) velocities and confirmed that the lack of rotation persists up to shock pressures of 70~GPa throughout the crystal.

It was shown in Sec.~\ref{sec:001Cu} that the typical shock-induced rotation suffered by $[001]$ Cu increased markedly when the crystals were pretilted with respect to the compression direction. One might wonder whether $[111]$ Cu behaves similarly, and whether the $[111]$ direction only appears to be stable because we are shocking along that direction exactly. There is in fact a modicum of truth in this statement. In Fig.~\ref{fig:111CuDeflectionScan}, we show pole plots for the $(111)$ planes for a set of five crystals all shocked to 70~GPa, but rotated around the $[1\bar{1}0]$ direction by angles between $0^\circ$ and $4^\circ$ before compression. As with $[001]$ Cu, perturbing the crystal's alignment favors slip on certain systems over others; in this case, slip on plane $(11\bar{1})$ (plane b) is favored, causing the crystal to rotate further about the $[1\bar{1}0]$ axis. However, the extent of this rotation is relatively small -- while $[001]$ Cu shocked to 30~GPa rotates by $4^\circ$ when pretilted by as a little as $1^\circ$, $[111]$ Cu shocked to the considerably greater pressure of 70~GPa and pretilted by some $4^\circ$ rotates by only $2.5^\circ$. Even when encouraged, then, $[111]$ Cu is `reluctant' to rotate when shock compressed.

To understand why $[111]$ Cu does not rotate under shock compression, we must explain why activity on the three slip planes is so similar. Why does $[111]$ Cu not deform on just two planes, as $[001]$ Cu does? We believe the basic difference between $[001]$ and $[111]$ Cu can be explained by considering the locus of stress and strain states available to each crystal under the highly restrictive constraints imposed by uniaxial compression conditions. We intend to show that $[111]$ Cu \emph{must} deform on at least three slip planes in order to reach a relaxed (i.e.,\ mechanically stable) state. We will first present a short strain-centric argument that captures the essential physics, before showing that without activating a third slip plane, $[111]$ Cu cannot access any stable states if it is laterally confined.

In Sec.~\ref{sec:001Cu}, we calculated the locus of elastic strain states available to $[001]$ Cu undergoing perfect duplex slip subject to a total strain $F = \text{diag}(1,1,v)$, arguing that the true state assumed by the crystal would be that for which the effective strain $\varepsilon_{\text{vms}}^e$ was minimal. In Fig.~\ref{fig:locicomparison}, we show explicitly how $\varepsilon_{\text{vms}}^e$ varies as a function of slip activity for $[001]$ Cu compressed by $25\%$. We see that by undergoing double slip, $[001]$ Cu is able to reduce the effective strain to just 20\% of its initial value. That is, the orientation of the slip planes and slip directions in $[001]$ Cu is such that double slip alone can take the crystal close to a hydrostatic strain state. Indeed, the pretilted MD simulations shown in Sec.~\ref{sec:001Cu} demonstrated that the finite residual shear strains left by double slip are small enough that they can be tolerated by a $[001]$ Cu crystal. We can construct a similar locus for $[111]$ Cu by assuming that it undergoes an analogous duplex-slip deformation mode (e.g.\ by slipping equally on systems c$2$, c$\bar{3}$, d$3$, and d$\bar{1}$). The resultant curve, also plotted in Fig.~\ref{fig:locicomparison}, shows that the effective strain in $[111]$ Cu can fall by no more than 40\% if only planes c and d are active -- the state of zero deviatoric elastic strain remains far beyond the crystal's reach. In other words, the double slip mode is efficient at relaxing shear strains in $[001]$ Cu, but is largely ineffectual in $[111]$ Cu.

To solidify this argument, we can quantify the inability of conjugate slip to relieve the shock-induced shear \emph{stresses} in $[111]$ Cu. Our goal is to establish whether there exists a mechanical state available to a $[111]$ Cu crystal slipping on just two $\{111\}$ planes for which slip on additional slip systems is not required. To do so, we first set up four slip parameters $\gamma_1, \gamma_2, \gamma_3$, and $\gamma_4$, which quantify the amount of slip on systems c$2$, c$\bar{3}$, d$3$, and d$\bar{1}$, respectively. Via Eq.~(\ref{eq:plasticdefpoly}), these slip parameters determine the plastic deformation gradient
    \begin{equation}
        \label{eq:plasticdeffour}
        F^p = I + \sum_{\alpha=1}^4 \gamma_\alpha(\mathbf{m}_\alpha \otimes \mathbf{n}_\alpha)
    \end{equation}
to leading order, which, when combined with Eqs.~(\ref{eq:elastoplastic},\ref{eq:totaldef}), gives the elastic deformation gradient $F^e$, from which we extract the pure strains using the polar decomposition in Eq.~(\ref{eq:decomp}). By expanding the elastic strains around a hydrostatic strain state of the same volume, we can estimate the Cauchy stresses $\sigma_{ij}$ using high-pressure elastic constants $C_{ijkl}$ obtained from first-principles calculations:
    \begin{equation}
        \label{eq:Cijkl}
        C_{ijkl}=\frac{\partial\sigma_{ij}}{\partial\varepsilon_{jk}^e},\quad i,j,k,l=x,y,z.
    \end{equation}
The resolved shear stress acting on each of the 12 fcc full slip systems can then be calculated using the usual resolution of the Cauchy stress tensor:
    \begin{equation}
        \label{eq:stressresolution}
        \tau_\alpha = \sum_{i,j} \sigma_{ij}[\mathbf{m}_\alpha]_i[\mathbf{n}_\alpha]_j.
    \end{equation}
From this set, we extract the greatest resolved shear stress acting on any of the 12 slip systems, $\tau_{\text{max}}$. These calculations are then embedded into an optimization routine that allows us to explore `slip space', and identify the combination of slip activities $\{\gamma_i\}$ that yields the smallest accessible value of $\tau_{\text{max}}$. In this way, we can locate the most mechanically stable state available to the crystal under the combined constraints of lateral confinement and slip on two planes only.

We carried out this optimization with a covariance matrix adaptation evolution strategy (CMA-ES) routine for a $[111]$ Cu crystal compressed by 18\% (the same total compression measured in the Milathianaki experiment), using elastic constants calculated using density-functional theory (DFT) by Li-Gang and Jing \cite{Li-Gang2010}. While these first-principles calculations capture the density dependence of the elastic moduli, we should note that they account for neither the thermal softening nor the work hardening that occurs during dynamic compression. The optimization found that the mechanical state for which the maximum resolved shear stress acting on any of the 12 fcc slip systems was minimized was accessed by slip activities $\gamma_1 = \gamma_4 = 0.03$ and $\gamma_2 = \gamma_3 = 0.20$ (i.e.,\ the slip was concentrated largely onto the shared slip direction common to the two active slip planes). The maximum shear stress for this most stable state was $\tau_{\text{max}} = 2.9$~GPa. Citing again the strength measurement performed by Murphy \emph{et.\  al.}\ \cite{Murphy2010}, who found that the typical strength of copper at $(28\pm2)\%$ compression was just $(1.1\pm0.2)$~GPa, we suggest a residual shear stress of 3~GPa could not be supported by copper at 18\% compression. This is to say that if $[111]$ Cu shocked to this pressure is permitted only to slip on two $\{111\}$ slip planes, there are no mechanically stable states available to it. It necessarily follows that the third slip plane must become operative, to relieve the components of shear stress that the first two slip planes cannot. As shown in the pole plots in Fig.~\ref{fig:111CuDeflectionScan}, the rotation brought about by the third slip plane directly opposes that caused by the first two. Of course, while this argument does not explain exactly \emph{how much} slip is required of the third plane, it does provide some intuition as to how stabilization of the $[111]$ direction is achieved in MD.

We turn now to the more important question of whether the experimental data support the prediction that the $[111]$ direction is stable to uniaxial compression. As previously noted, the scattering geometry of Milathianaki \emph{et.\ al.}'s experiment \cite{Milathianaki2013} is such that if the crystal rotates enough, there will be a marked reduction in the scattering intensity from the plastically relaxed region of the crystal. In principle, this sensitivity to slip-induced rotation provides a means of discriminating between conjugate slip (of the kind suffered by $[001]$ Cu under shock) and fully symmetric slip (as observed in our $[111]$ Cu MD simulations). However, it transpires that the rotation associated with even the relatively asymmetric conjugate-slip state is small enough that appreciable diffraction can still take place, despite the experimental samples' narrow texture width.

To demonstrate this, we applied a simple diffraction model that predicts the signal generated by a fiber-textured fcc polycrystal with a prescribed, grain-specific elastic strain state. In brief, the model first generates a set of $\{111\}$ scattering vectors consistent with an ambient $[111]$ fiber-textured copper polycrystal. The deviation of each grain's orientation from the ideal texture orientation was drawn from a pseudo-Voigt distribution with a $2.9^\circ$ half-width at half-maximum (HWHM), which represents reasonably well the crystallographic texture of pre-characterized experimental samples. Given a total compression $V/V_0$ and a plastic deformation mode (i.e.\ the ratios between the slip activities $\{\gamma_i\}$), the deviatoric-strain-minimizing deformation gradient $F^e$ is calculated, and its reciprocal-space version $[(F^e)^T]^{-1}$ is applied to the $\{111\}$ scattering vectors, yielding the deformed reciprocal lattice. The scattering vectors are then convolved with a Gaussian `shape function' that gives them a finite extent and is responsible for the linewidth of the Bragg peaks. Finally, the intensity of the reciprocal lattice is calculated on the surface of the Ewald sphere and azimuthally integrated to give a one-dimensional diffraction pattern. Further details about the structure and validation of this diffraction model are located in the Supplemental Material \cite{supplemental}.

\begin{figure}[t]
    \includegraphics{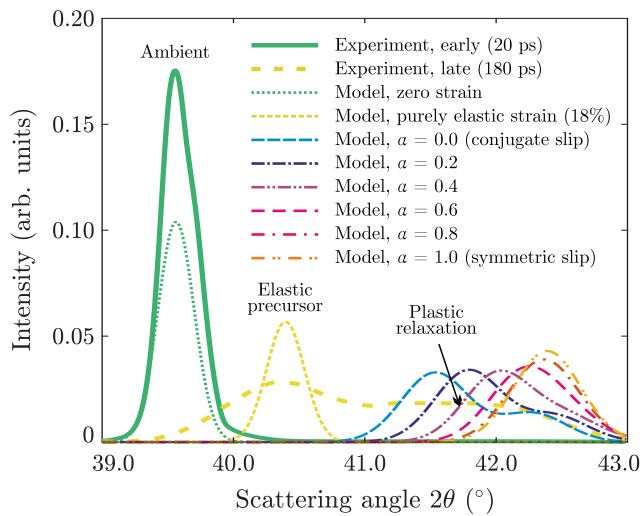}
    \caption{Comparison of early- and late-time experimental diffraction signals from dynamically compressed, $[111]$ fiber-textured copper polycrystals loaded to 73~GPa peak normal stress by Milathianaki \emph{et.\ al.}\  \cite{Milathianaki2013} with simulated diffraction signals from crystals in various idealized deformation states. The intensity of the simulated diffraction has been scaled for ease of comparison with the data. Included are simulated signals from ambient material, from material in an elastic precursor wave at 18\% compression, and for plastically relaxed material deforming on between two and three slip planes; the impurity parameter $\alpha$ expresses how much the third slip plane participates in the relaxation process.}
    \label{fig:111CuDiffraction}
\end{figure}

We used this model to measure the attenuation of the diffraction signal caused by slip-induced rotation in order to assess whether duplex slip is ruled out by the experimental data. In Fig.~\ref{fig:111CuDiffraction}, we show a comparison of the experimental data of Milathianaki \emph{et.\ al.}\ with modeled diffraction signals from selected idealized deformation modes. The early-time data, taken just 20~ps after compression begins, is dominated by a single peak at $2\theta = 39.6^\circ$ generated by the largely unstrained sample. The late-time data is distributed over a considerably larger range of scattering angles, coming as it does from a dynamically compressed sample containing a depth-dependent distribution of strain states. Intensity collects first around a broad peak centered at $40.4^\circ$ due to the presence of an elastic precursor wave, which, as noted in the original study, is consistent with a uniaxial elastic strain of 18\%. A wide plateau also forms between $41^\circ$ and $43^\circ$, created by material spanning the gamut of deformation states between purely elastic strain and almost complete plastic relaxation. In the original study, the diffraction signal could be accounted for very well by assuming that the samples plastically deformed much as our simulated crystals did, which is to say, with symmetric activation of all three slip planes. We now wish to determine whether the signal can also be explained by an asymmetric deformation mode like double slip.

We show in Fig.~\ref{fig:111CuDiffraction} a series of synthetic diffraction signals from plastically relaxed material at 18\% total compression. Each curve corresponds to a different value of an impurity parameter $\alpha$ similar to that used in Sec.~\ref{sec:001Cu}, where $\alpha = 0$ denotes ideal conjugate slip (specifically, equal slip on systems c$2$, c$\bar{3}$, d$3$, and d$\bar{3}$) and where $\alpha = 1$ corresponds to fully symmetric slip (i.e., equal activation of not only c$2$, c$\bar{3}$, d$3$, and d$\bar{3}$, but also b$1$ and b$\bar{2}$, as seen in the MD). We observe that as $\alpha$ decreases, the diffraction signal from the plastically relaxed portion of the crystal undergoes three changes. First, we see that the signal actually separates into two overlapping peaks of different heights. This asymmetric splitting stems from the shear strains present in the double-slipped state, and we will not comment further upon it here. Second, we observe that as more and more slip is concentrated onto just two planes, the scattering per unit volume does \emph{not} decrease markedly. In fact, even the most rotated material (with $\alpha = 0$) scatters more strongly than the material in the elastic precursor. That is, the rotation brought about by double slip is not enough to significantly reduce the diffraction intensity despite the samples' narrow texture width. While the total rotation attending perfect conjugate slip is $3.8^\circ$, the greatest deflection suffered by any of the $\{111\}$ scattering vectors when resolved perpendicular to the Ewald sphere [i.e., the angle $\omega$ depicted in Fig.~\hyperref[fig:polanyi]{16(e)}] in only $1.8^\circ$. Since this is less than the texture's HWHM of $2.9^\circ$, the double-slip diffraction signal remains strong enough to account for the higher-angle data -- conjugate slip cannot be discounted on the basis of its intensity alone.

The third property of the diffraction signal to change is the average scattering angle. As the plastic deformation mode transitions from ideal double slip to fully symmetric slip, $\langle2\theta\rangle$ increases monotonically until hitting the upper limit represented by a fully hydrostatic strain state ($42.4^\circ$ for 18\% compression). We would argue that we \emph{can} rule out perfect double slip $(\alpha = 0)$ due to its diffraction signal being significantly attenuated above $\sim42.5^\circ$ -- were the crystal to deform on only two slip planes, the plateau could not extend to $43^\circ$. However, for values of $\alpha$ above 0.6, the differences in peak shape, height, and position become marginal, such that any impurity above $\alpha>0.6$ could plausibly account for the experimental data, given appropriate density profiles. In brief, we contend that the experiment rules out perfect double slip (as expected from the mechanical instability of the associated strain state) but cannot discriminate between fully symmetric slip and imperfect conjugate slip. To offer this discriminating power, an alternative scattering geometry is needed, as we will discuss in Sec.~\ref{sec:discussion}.

In summary, we performed simulations of monocrystalline copper shocked along $[111]$ and observed that slip was distributed equally between three $\{111\}$ slip planes, leading to zero net rotation of the crystal structure. We argued that all three slip planes must eventually become active to some extent in order that the crystal can reach a mechanically stable state, thus stabilizing the $[111]$ direction to uniaxial compression. By comparing synthetic diffraction patterns from crystals plastically deforming on two or three slip planes with experimental data taken by Milathianaki \emph{et.\ al.}\ \cite{Milathianaki2013} from dynamically compressed copper polycrystals, we found that the hypothesis that copper is stable to compression along $[111]$ was not contradicted by the data. However, the limited discriminating power offered by the scattering geometry meant the data was also consistent with marginally asymmetric slip.

\section{\label{sec:discussion} Discussion}

We have revisited three x-ray diffraction experiments performed on [101] Ta (Wehrenberg \emph{et.\ al.}\ \cite{Wehrenberg2017}), [001] Cu (Suggit \emph{et.\ al.}\ \cite{Suggit2012}), and [111] Cu (Milathianaki \emph{et.\ al.}\ \cite{Milathianaki2013}) with the intention of deepening our understanding of the plasticity mechanisms activated in each by rapid uniaxial compression. We have used large-scale molecular dynamics simulations to predict which combinations of crystal planes accommodate the plastic strain, and, in all three cases, the simulations appear to be consistent with their corresponding experiment. Our results underscore how important dedicated texture studies like those performed in the above-mentioned experiments are for advancing our knowledge of high strain-rate plasticity, as well as highlighting the invaluable role classical MD simulations can play as interpretative and predictive tools for such experiments. We close with a few remarks about natural directions for future work following this study.

Much of this work focuses on explaining why certain crystallographic directions are stable or unstable to compression by accounting for the plasticity mechanisms that become active under shock. However, we can offer only limited insights into why those mechanisms become active in the proportion that they do. Taking [001] Cu as an example, the tendency of certain $\langle112\rangle \{111\}$ stacking faults to dominate could be broadly understood by comparing the resolved shear stresses acting upon them (essentially applying Schmid's law). However, we can only speculate as to why – assuming our simple rotation model is correct – the dominant and contaminant slip systems are activated in the particular ratio $\sim 80:20$.

To understand the rapid plastic deformation process on this quantitative level requires analysis of the kind performed by Avraam \emph{et.\ al.}\ \cite{Avraam2021}. There, the authors used a dislocation-aware, rate-dependent strength model for tantalum to calculate the instantaneous slip rates $\{\dot{\gamma}_\alpha\}$ on its twelve $\langle111\rangle\{112\}$ systems under shock-loading conditions. By choosing the homogeneous nucleation rate and threshold appropriately, they could tune the slip rates so as to reproduce the correct lattice rotation measured by Wehrenberg \emph{et.\ al.}\ along the Hugoniot. The authors found that the ratio between the dominant slip systems activities was pressure-sensitive, and could explain the sharp jump in the rotation angle above a 26~GPa threshold. We believe an analogous model for dynamically compressed copper could exploit the rotation measurements of Suggit \emph{et.\ al.}\ in a similar way to constrain the dislocation kinetics controlling plastic deformation in copper under extreme loading conditions.

Perhaps the most interesting prediction from our simulations is that copper uniaxially shocked along [111] experiences no change to its crystallographic texture. Kinematic analysis revealed that this stability comes about because plasticity is shared equally among the active slip systems, meaning the sample retains its original three-fold symmetry. This behavior contrasts with copper’s response when compressed along [001], where a subset of the symmetrically equivalent plasticity mechanisms become active, breaking the sample’s four-fold symmetry and allowing the crystal structure to rotate.

While the dynamic compression data of Milathianaki \emph{et.\ al.}\ does not disprove the prediction that [111] Cu is stable to uniaxial compression, it does not unambiguously confirm it, either. Its inability to resolve small texture changes is a consequence of its scattering geometry, in which the quasimonochromatic x-ray beam is launched parallel to the fiber axis. In this configuration, texture changes cause a reduction in the integrated scattering intensity, the resolution of which can become impractical if the slip-induced rotation is smaller than the texture width (as seems to be the case here). If instead the x-rays impinge on the target at non-normal incidence, the symmetry of the diffraction pattern is broken, and continuous texture changes manifest as broadening and eventual splitting of discrete diffraction peaks. It was this property of the scattering geometry that allowed Wehrenberg \emph{et.\ al.}\ to measure shock-driven orientation changes in nanocrystalline tantalum with sub-degree precision. An experiment similar to that of Milathianaki \emph{et.\ al.}\ focusing on copper shocked along $[111]$, but probed with x-rays at an angle to the target normal, would therefore be highly valuable, and would allow us to confirm or refute the predictions of our simulations.

\section{\label{sec:conclusion} Conclusion}

We have carried out large-scale molecular dynamics simulations of single-crystal tantalum and copper under uniaxial shock-compression conditions, and compared their texture evolution with that of real dynamically compressed crystals measured using ultrafast x-ray diffraction. In agreement with the results of Wehrenberg \emph{et.\ al.}\ \cite{Wehrenberg2017}, we found that tantalum shocked along $[101]$ was unstable to shock compression, and suffers slip-induced rotation about its local $[10\bar{1}]$ axis. Kinematic analysis of the simulated crystals showed that this rotation arises thanks to asymmetric activation of the crystal's two primary slip systems combined with completely symmetric activation of its secondary slip systems. We found that copper shocked along $[001]$ is also unstable to uniaxial compression, and plastically deforms on pairs of slip planes that cause rotation about the $\langle100\rangle$ directions, as observed in the experiment of Suggit \emph{et.\ al.}\ \cite{Suggit2012}. We used a simple rotation model to show that the data are best-explained by imperfect conjugate slip, in which approximately 80\% of the plasticity is mediated by two dominant slip planes, the remaining 20\% being taken up by `contaminant' slip planes. We also studied copper shocked along $[111]$, which remained stable under compression due to equal activation of its six dominant slip systems. We found that although this lack of rotation was consistent with the dynamic compression experiment of Milathianaki \emph{et.\ al.}\ \cite{Milathianaki2013}, the data could also be explained by a plastic response mediated by marginally asymmetric slip. These results highlight the crucial role competition between plasticity mechanisms plays in controlling the texture evolution of crystalline matter under uniaxial compression, in addition to the importance of time-resolved x-ray diffraction experiments in developing our understanding of the underlying physics.

\begin{acknowledgments}
The authors would like to thank P.~Avraam, E.~Floyd, and D.~McGonegle for valuable discussions in the early stages of this work. P.~G.~H.\ thanks Livermore Computing for providing computing resources. Both P.~G.~H.\ and J.~S.~W.\ gratefully acknowledge the support of AWE via the Oxford Centre for High Energy Density Science (OxCHEDS). J.~S.~W.\ further acknowledges support from EPSRC under Grant No.\ EP/S025065/1.
\end{acknowledgments}

%

\end{document}


\title{Supplemental Material: Slip competition and rotation suppression in tantalum and copper during dynamic uniaxial compression}

\author{P.~G.~Heighway}
\email{patrick.heighway@physics.ox.ac.uk}
\affiliation{Department of Physics, Clarendon Laboratory, University of Oxford, Parks Road, Oxford, OX1 3PU, United Kingdom}
\author{J.~S.~Wark}
\affiliation{Department of Physics, Clarendon Laboratory, University of Oxford, Parks Road, Oxford, OX1 3PU, United Kingdom}

\date{\today}

\pacs{}
\maketitle

\section*{Diffraction model}

Here we set out the structure of the diffraction model we used to simulate elastic x-ray scattering from $[111]$ fiber-textured copper polycrystals during dynamic compression. Our model takes the elastic strains and rotation resulting from predetermined plastic deformation modes and uses them to predict the attendant changes to the position, intensity, and structure of the $\{111\}$ diffraction peak. Its purpose is to constrain the plasticity mechanisms operative in the copper polycrystals examined by Milathianaki \emph{et.\ al.}\ \cite{Milathianaki2013} using the time-resolved x-ray diffraction patterns they obtained from their targets at peak normal stresses of $\sim70$~GPa.

The calculation begins with the creation of a set of grains whose distribution of orientations is consistent with a fiber texture. Each grain is generated with a default orientation in which the $[1\bar{1}0]$, $[11\bar{2}]$, and $[111]$ directions are aligned with the $x$-, $y$-, and $z$-axes, respectively, where $z$ represents the fiber axis. The grain is randomly assigned Euler angles $(\alpha,\beta,\gamma)$ and actively rotated with the operator $R$ generated by the $z$-$x$-$z$ sequence:
    \begin{equation}
        R = R_z(\alpha)R_x(\beta)R_z(\gamma).
    \end{equation}
To be consistent with a fiber texture, angles $\alpha$ and $\gamma$ are uniformly sampled from the interval $[0,2\pi)$, while $\beta$ (which measures the deviation of the grain's orientation from the ideal fiber orientation) is distributed according to the probability density function
    \begin{align}
        P(\beta) &= \sin\beta\times f(\beta); \\
        f(\beta) &= \frac{\eta}{1+(\beta/\delta\beta)^2} + (1-\eta)\exp\left[-\ln 2 \left(\frac{\beta}{\delta\beta}\right)^2\right].
    \end{align}
The function $f$ approximates a Voigt profile that ensures the probability of a given grain deviating by an angle $\beta$ decreases monotonically over a characteristic angle $\delta\beta$ we refer to as the `texture width'; specifically, $\delta\beta$ is the half-width at half-maximum (HWHM) of the distribution. The parameter $\eta$ is the relative weighting of the Lorentzian and Gaussian contributions to the profile. The functional form of $f$ is purely phenomenological -- the choice was not motivated by any particular knowledge of the physics of the grain formation process. The multiplicative prefactor of $\sin\beta$ in the expression for $P(\beta)$ is a density-of-states--like factor that ensures grains are not oversampled near the poles (at $\beta=0$ and $\pi$).

\begin{figure}[tb]
    \includegraphics{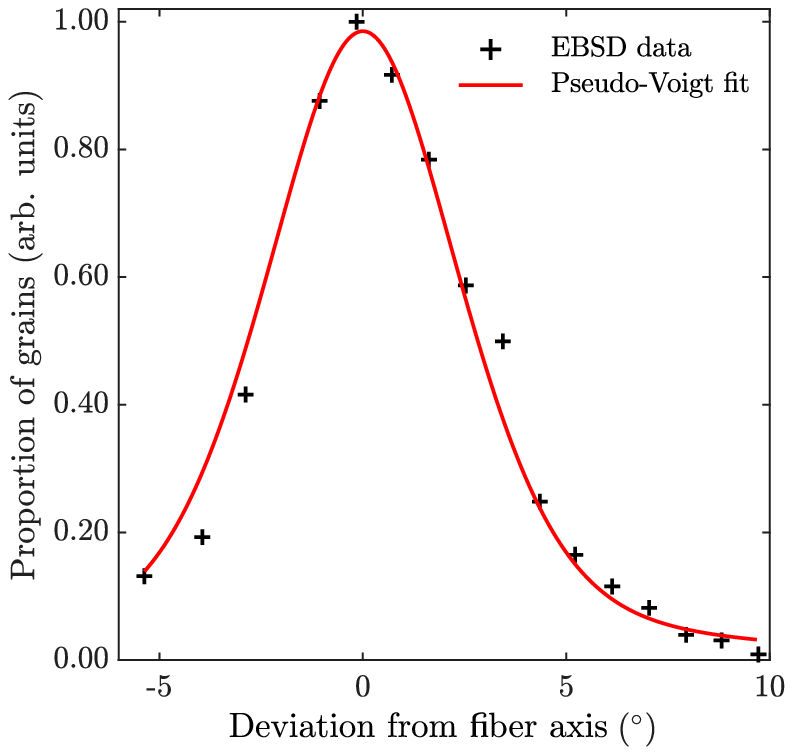}
    \caption{Distribution of grain deviations from the $[111]$ fiber axis for polycrystalline copper films used in the dynamic-compression experiment of Milathianaki \emph{et.\ al.}\ \cite{Milathianaki2013}, as measured using electron backscatter diffraction (EBSD). Shown also is the phenomenological, Voigt-like fit used to generate grain distributions in our diffraction model.}
    \label{fig:EBSD}
\end{figure}

\begin{figure}[tb]
    \includegraphics{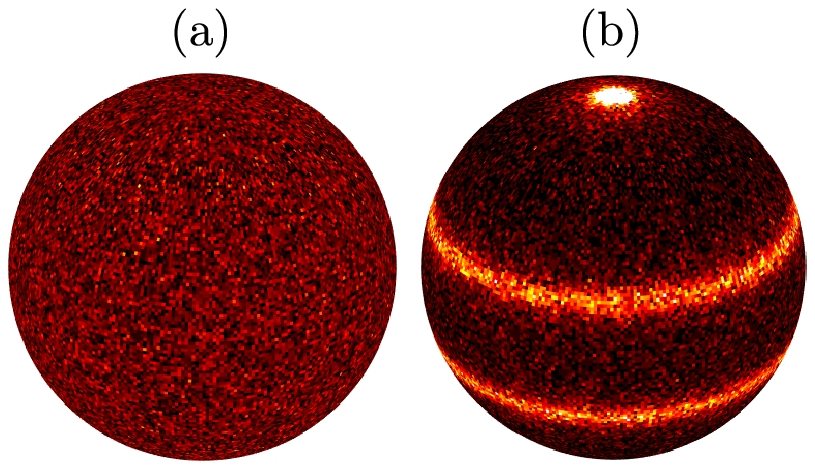}
    \caption{Representative distributions of $\{111\}$ plane normals in polycrystals generated in our diffraction model, pictured as heatmaps on the unit sphere. (a) Untextured sample, generated using an extremely large texture width. (b) Fiber-textured sample representative of the experiment, generated with a texture width of $\delta\beta=2.9^\circ$.}
    \label{fig:jovian}
\end{figure}

Parameters $\delta\beta$ and $\eta$ were obtained by fitting the function $f(\beta)$ to an experimental measurement of the distribution of grain deviations obtained via electron backscatter diffraction (EBSD). The fit to the texture data, shown in Fig.~\ref{fig:EBSD}, yields a texture width of $\delta\beta = 2.9^\circ$ and a Lorentz-Gauss weighting of $\eta = 0.40$. With these parameters, $P(\beta)$ generates the distribution of $\{111\}$ plane normals shown on the surface of the unit sphere in Fig.~\hyperref[fig:jovian]{2(b)}. We see intense `caps' at the poles of the sphere that come from the tightly clustered $(111)$ and $(\bar{1}\bar{1}\bar{1})$ planes, in addition to characteristic `tropics' at $\pm19.5^\circ$ latitude from the $\{\bar{1}11\}$ and $\{\bar{1}\bar{1}1\}$ planes. We also show for reference the distribution of $\{111\}$ plane normals generated for an extremely large value of $\delta\beta$ in Fig.~\hyperref[fig:jovian]{2(a)}, for which we recover the uniform distribution of orientations expected of an untextured, powderlike sample.

The default population of our simulated polycrystals is $N_G = 14\,000$ grains. The samples used by Milathianaki \emph{et.\ al.}\ were determined by EBSD to have an average grain size of $400$~nm \cite{Milathianaki2013}. Our value of $N_G$ is based on an estimation of the number of crystallites in a micron-thick target that would be illuminated by a $30\times30$~\textmu m\textsuperscript{2} x-ray spot. We assume all grains are of the same size (i.e., each has the same scattering strength).

The next step is to distort the scattering plane normals in a manner consistent with the plastic deformation suffered by the polycrystal. The key assumption underpinning this kinematic part of the model is that once the polycrystal has plastically relaxed, every grain has the same elastic deformation gradient \emph{in its own crystal frame}. That is to say that every crystallite suffers the same elastic strain along its local $[111]$ axis, undergoes rotation about the same crystallographic axis, and so on. We make this homogenization assumption with the understanding that, in reality, both grain-grain interactions and the deviation of each grain from the $[111]$ fiber direction will cause the form of $F^e$ to vary slightly from grain to grain. Under the action of the grain-specific, real-space elastic deformation gradient $F^e$, each crystal plane normal $\mathbf{n}_i$ transforms as
    \begin{equation}
        \mathbf{n}_i \to [(F^e)^T]^{-1}\mathbf{n}_i.
    \end{equation}

Having obtained the set of distorted crystals planes, we can then generate the full reciprocal-space intensity of the crystal $I(\mathbf{k})$. We first identify the set of all ideal scattering vectors $\{\mathbf{G}_i\}$ associated with the crystal planes:
    \begin{equation}
        \mathbf{G}_i = \sqrt{3}\left(\frac{2\pi}{a_0}\right)\mathbf{n}_i.
    \end{equation}
If all grains were defect-free and of infinite dimensions, $I(\mathbf{k})$ would simply be a set of delta-functions centered on these scattering vectors:
    \begin{equation}
        I_\infty(\mathbf{k}) = \sum_i \delta(\mathbf{k} - \mathbf{G}_i).
    \end{equation}
To encode broadening effects, we convolve this signal with a three-dimensional Gaussian shape function $I_s$:
    \begin{align}
        I(\mathbf{k}) &= \int d^3\mathbf{k}'\,I_s(\mathbf{k}')I_{\infty}(\mathbf{k} - \mathbf{k}'); \\
        I_s(\mathbf{k}) &= \frac{1}{\left(\sqrt{2\pi}\sigma\right)^3} \exp\left(-\frac{\mathbf{k}^2}{2\sigma^2}\right).
    \end{align}
The width $\sigma$ encodes the linewidth of the resulting diffraction peaks. For simulation of signals from crystals that have not plastically deformed, we simply choose the width $\sigma$ such that the linewidths match that of the ambient diffraction signal measured by Milathianaki \emph{et.\ al.} For simulation of plastically relaxed material, we match instead the linewidth predicted by the model of Bragg \cite{Bragg1949}, in which broadening results from the microscopic strain inhomogeneities surrounding dislocations. Following Foster \emph{et.\ al.}\ \cite{Foster2017}, we estimate the the broadening of the $\{111\}$ peaks to scale with the dislocation density $\rho$ according to
    \begin{equation}
        \Delta(2\theta)_{111} = \frac{1}{\sqrt{2}\cos\theta_{111}}\left(\lambda\rho^{\frac{1}{2}}\right),
    \end{equation}
where $\theta_{111}$ is the central scattering angle and $\lambda=1.41$~\AA\ is the x-ray wavelength. Using typical values of the dislocation density reached at these strain rates in dislocation dynamics simulations \cite{Shehadeh2006} ($\rho\sim10^{12}$~cm\textsuperscript{$-2$}), we estimate a broadening of approximately $0.6^\circ$. This is the linewidth used in the diffraction patterns of plastically relaxed material in Fig.~21 of the Main Article.

To generate the full two-dimensional scattering pattern, the reciprocal-space intensity $I(\mathbf{k})$ is sampled on the Ewald sphere. Given the collinearity of the x-ray direction and fibre axis in Milathianaki \emph{et.\ al.}'s experiment, Ewald's sphere is the locus of $k$-points satisfying
    \begin{equation}
        \left(\mathbf{k} + \frac{2\pi}{\lambda}\mathbf{e}_z\right)^2 = \left(\frac{2\pi}{\lambda}\right)^2.
    \end{equation}
The intensity is then azimuthally integrated around the Ewald sphere to give one-dimensional diffraction patterns to compare with the experimental data.

Since we are concerned with only a small range $(\sim 5^\circ)$ of scattering angles, we neglect the $\theta$-dependent intensity variations resulting from both the atomic form factor and polarization effects. We also ignore self-absorption, because the attenuation length for 8.8~keV x-rays in copper (28~\textmu m \cite{Milathianaki2013}) is far greater than the target thickness (1~\textmu m).

\section*{Test cases}

\begin{figure}[t]
    \includegraphics{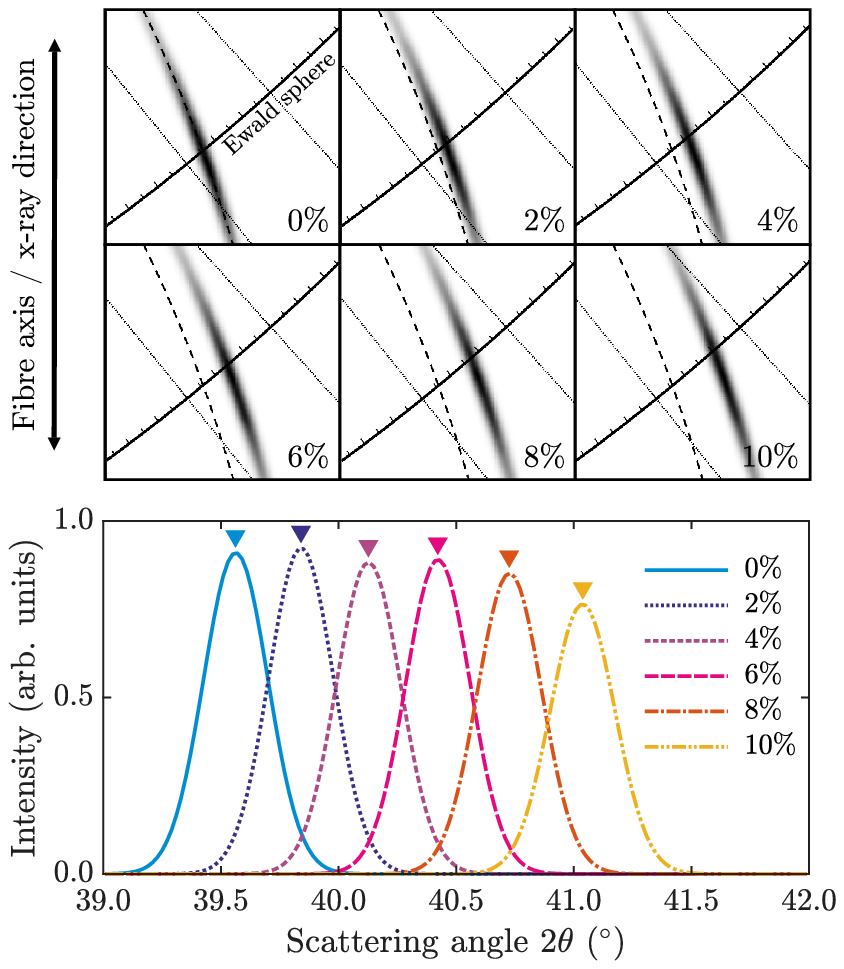}
    \caption{Modeling of elastic x-ray diffraction from $[111]$ fiber-textured copper probed with an 8.8~keV beam traveling parallel to its fiber axis for hydrostatic elastic strains ranging between 0\% and 10\%. Top: visualization of the azimuthally integrated reciprocal-space intensity of the polycrystal in the vicinity of the actively scattering $\{111\}$ planes. Solid line represents a cross-section of the Ewald sphere, the dashed line that of the $\{111\}$ Polanyi surface. Dotted lines intersect the Ewald sphere at equivalent scattering angles of $2\theta = 39^\circ$ and $42^\circ$. Bottom: azimuthally integrated diffraction patterns. Inverted triangles mark the ideal center of each peak predicted by Eq.~(\ref{eq:hydro}) from Bragg's law.}
    \label{fig:hydrostatic}
\end{figure}

To test the accuracy of our model, we examined three simple test cases. The first test involves hydrostatically compressing the crystal structure with diagonal the elastic deformation gradient
    \begin{equation}
        F^e_{\text{hydro}} = (1 + \varepsilon^e)I.
    \end{equation}
Bragg's law tells us that if all atomic plane spacings are identically scaled by factor $(1+\varepsilon^e)$, the new Bragg angle~$\theta$ of the family of scattering planes with original Bragg angle $\theta_0$ satisfies
    \begin{equation}
    \label{eq:hydro}
    (1+\varepsilon^e)\sin\theta = \sin\theta_0.
    \end{equation}
We generated diffraction patterns for values of $\varepsilon$ chosen so as to give volumetric compressions between 0\% and 10\% in steps of 2\%, and compared the positions of the peaks with those predicted by Eq.~(\ref{eq:hydro}). As shown by Fig.~\ref{fig:hydrostatic}, the Bragg peaks appear where they are expected to. We also notice a slight attenuation of the peak heights at higher values of compression, which is caused by the center of the `tropic' moving slightly away from the surface of the Ewald sphere, as pictured at the top of Fig.~\ref{fig:hydrostatic}. This intensity decrease is not perfectly monotonic thanks to the stochastic distribution of grain orientations.

\begin{figure}[t]
    \includegraphics{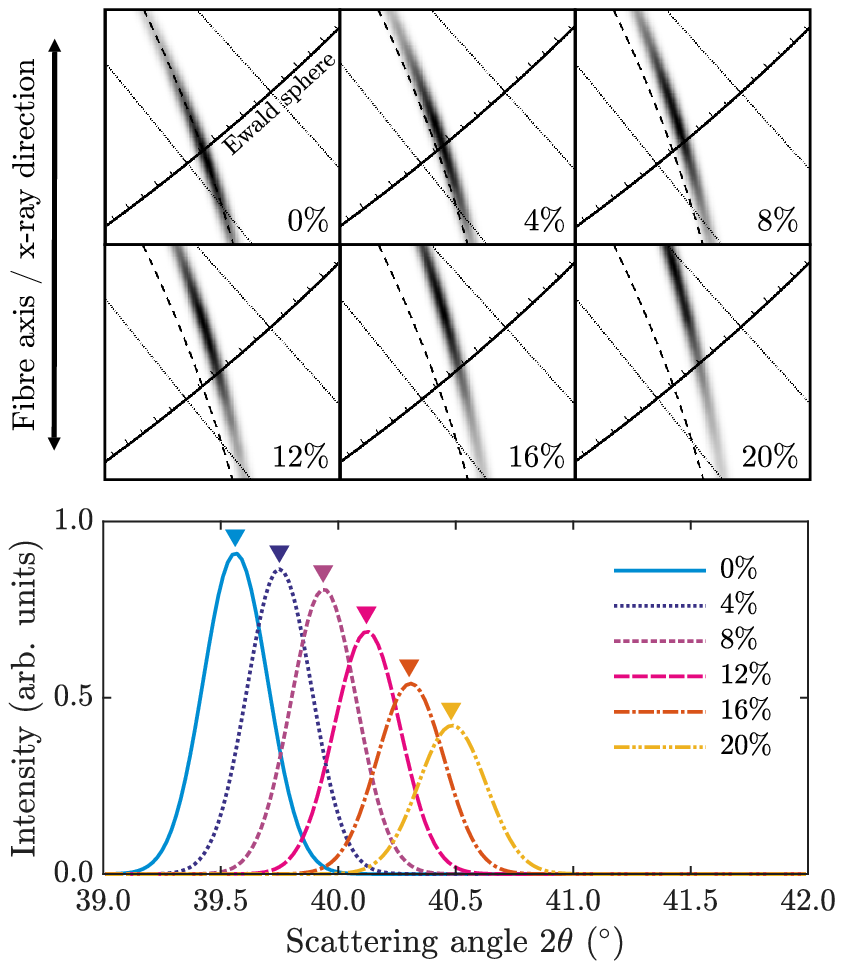}
    \caption{Modeling of elastic x-ray diffraction from $[111]$ fiber-textured copper probed with an 8.8~keV beam traveling parallel to its fiber axis for uniaxial elastic strains along the fiber axis ranging between 0\% and 20\%. Top: visualization of the azimuthally integrated reciprocal-space intensity of the polycrystal in the vicinity of the actively scattering $\{111\}$ planes. Solid line represents a cross-section of the Ewald sphere, the dashed line that of the $\{111\}$ Polanyi surface. Dotted lines intersect the Ewald sphere at equivalent scattering angles of $2\theta = 39^\circ$ and $42^\circ$. Bottom: azimuthally integrated diffraction patterns. Inverted triangles mark the ideal center of each peak predicted by Eq.~(\ref{eq:uniaxial}) from Higginbotham and McGonegle \cite{Higginbotham2014}. }
    \label{fig:uniaxial}
\end{figure}

The second test we ran involves elastic compression of the polycrystal along its fiber axis:
    \begin{equation}
        F^e_{\text{uniaxial}} = \begin{pmatrix}
        1 & 0 & 0 \\ 0 & 1 & 0 \\ 0 & 0 & 1 + \varepsilon^e_{zz}.
        \end{pmatrix}
    \end{equation}
Here, we assume that the elastic deformation gradient is identical everywhere in the `laboratory' frame (unlike the grain-specific deformation gradient we apply for plastically relaxed polycrystals). According to the model of Higginbotham and McGonegle \cite{Higginbotham2014}, the Bragg angle $\theta$ of a family of scattering planes with original Bragg angle $\theta_0$ subjected to purely elastic strain $\varepsilon_{zz}^e$ along $z$ satisfies
    \begin{equation}
        \label{eq:uniaxial}
        \left[(1+\varepsilon^e_{zz})^2 - 1\right]\sin^4\theta + \sin^2\theta = \sin^2\theta_0.
    \end{equation}
Fig.~\ref{fig:uniaxial} shows the variation of the $\{111\}$ diffraction peaks with uniaxial elastic strain $\varepsilon^e_{zz}$ up to $(-)20\%$. The peak positions are in accordance with Eq.~(\ref{eq:uniaxial}). We also draw the reader's attention to the fact that the peak intensity is significantly attenuated by the uniaxial strain, falling to around half of its ambient value once the strain crosses 18\%, the amount of compression measured by Milathianaki \emph{et.\ al.}\ \cite{Milathianaki2013}. This, as shown by Fig.~\ref{fig:uniaxial}, is a consequence of the uniaxial elastic strain pushing the majority of the narrow distribution of scattering plane normals `inside' the Ewald sphere, leaving only the wing of the distribution in the Bragg condition. In other words, material in an elastic precursor wave would diffract considerably more weakly than ambient material. We note that this texture-related attenuation effect did not appear to be factored into Milathianaki \emph{et.\ al.}'s reconstruction of the elastic strain profiles from the diffraction pattern.

\begin{figure}[t]
    \includegraphics{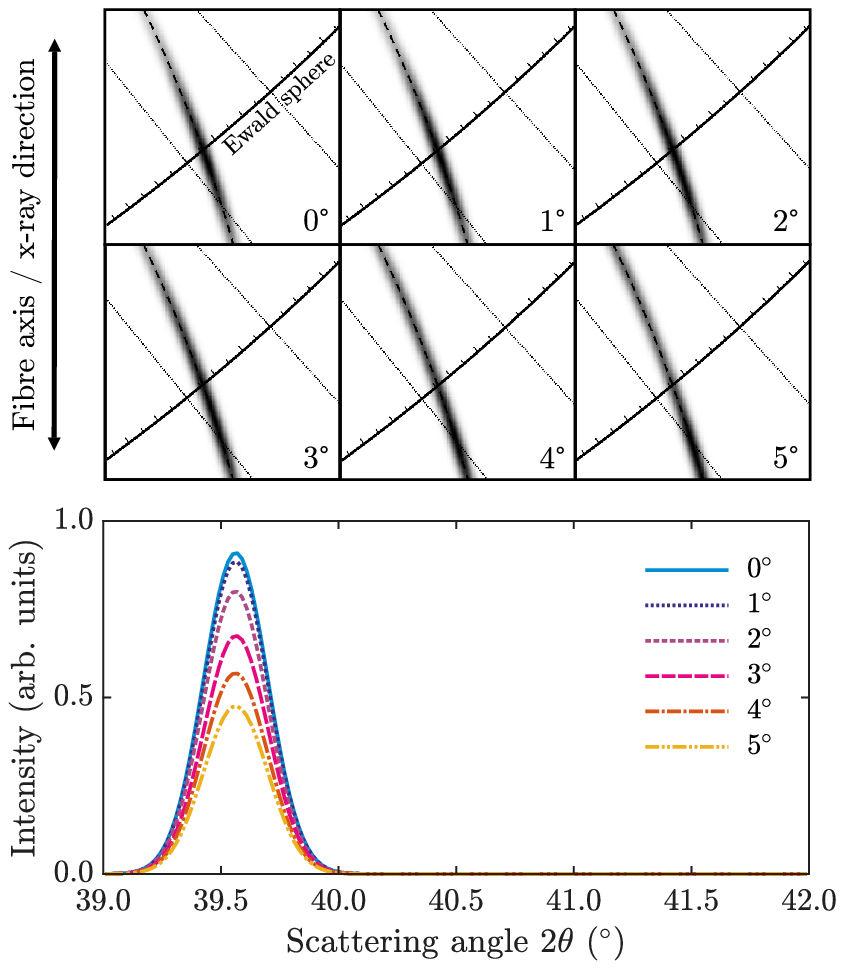}
    \caption{Modeling of elastic x-ray diffraction from $[111]$ fiber-textured copper probed with an 8.8~keV beam traveling parallel to its fiber axis for rotations around the local $[1\bar{1}0]$ axis by between $0^\circ$ and $5^\circ$. Top: visualization of the azimuthally integrated reciprocal-space intensity of the polycrystal in the vicinity of the actively scattering $\{111\}$ planes. Solid line represents a cross-section of the Ewald sphere, the dashed line that of the $\{111\}$ Polanyi surface. Dotted lines intersect the Ewald sphere at equivalent scattering angles of $2\theta = 39^\circ$ and $42^\circ$. Bottom: azimuthally integrated diffraction patterns.}
    \label{fig:rotation}
\end{figure}

The third test involves pure rotation of each grain about its local $[1\bar{1}0]$ axis. This is the axis around which each crystallite rotates if it undergoes equal amounts of slip on its $(\bar{1}11)$ and $(1\bar{1}1)$ planes -- this is the conjugate slip mode examined in the Main Article. For pure rotation, we do not expect any change in scattering angle, but the polycrystal's finite texture width means we do anticipate changes in intensity. In Fig.~\ref{fig:rotation}, we plot the $\{111\}$ scattering peak for rotations between $0^\circ$ and $5^\circ$. As expected, we observe that the peak intensity decreases monotonically with rotation due to the center of the distribution of $\{111\}$ planes moving out of the diffraction condition. Note that the distribution of scattering planes actually splits in two and migrates in `both directions' along the Polanyi surface. We find that around $4.5^\circ$ of rotation is required to halve the scattering intensity (for constant strains). This level of tolerance of the scattering to lattice rotation is consistent with our conclusion that the typical slip-induced rotation expected of $[111]$ Cu is insufficient to significantly attenuate the diffraction.

%